\documentclass[review,authoryear,12pt]{elsarticle}

\usepackage{graphicx}
\graphicspath{ {./Figures/} }
\usepackage[tight,footnotesize, nooneline]{subfigure}

\usepackage{adjustbox} 
\usepackage[hidelinks]{hyperref} 

\usepackage{amssymb}
\usepackage{amsthm}
\usepackage{amsmath}

\usepackage{lineno}
 
\usepackage{multirow}

\usepackage[flushleft]{threeparttable} 

\journal{arXiv}

\newcommand\copyrighttext{%
   \textcopyright\ 2018 IEEE.}

\begin{document}

\begin{frontmatter}



\title{Full modelling of high-intensity focused ultrasound and thermal heating in the kidney using realistic patient models}





\author[Affil1]{Visa Suomi \corref{cor1}}
\author[Affil2]{Jiri Jaros}
\author[Affil3]{Bradley Treeby}
\author[Affil1]{Robin Cleveland}
\address[Affil1]{Department of Engineering Science, University of Oxford, Parks Road, Oxford, OX1 3PJ, UK}
\address[Affil2]{Faculty of Information Technology, Brno University of Technology, Brno, Czech Republic}
\address[Affil3]{Department of Medical Physics and Biomedical Engineering, University College London, Wolfson House, 2-10 Stephenson Way, London, NW1 2HE, UK}
\cortext[cor1]{Corresponding Author: Visa Suomi, Institute of Biomedical Engineering, Old Road Campus Research Building, University of Oxford, Oxford OX3 7DQ, UK; Email, visa.suomi@oxon.org}

\begin{abstract}
\textit{Objective:} High-intensity focused ultrasound (HIFU) therapy can be used for non-invasive treatment of kidney (renal) cancer, but the clinical outcomes have been variable. In this study, the efficacy of renal HIFU therapy was studied using nonlinear acoustic and thermal simulations in three  patients. \textit{Methods:} The acoustic simulations were conducted with and without refraction in order to investigate its effect on the shape, size and pressure distribution at the focus. The values for the attenuation, sound speed, perfusion and thermal conductivity of the kidney were varied over the reported ranges to determine the effect of variability on heating. Furthermore, the phase aberration was studied in order to quantify the underlying phase shifts using a second order polynomial function. \textit{Results:} The ultrasound field intensity was found to drop on average 11.1 dB with refraction and 6.4 dB without refraction. Reflection at tissue interfaces was found to result in a loss less than 0.1 dB. Focal point splitting due to refraction significantly reduced the heating efficacy. Perfusion did not have a large effect on heating during short sonication durations. Small changes in temperature were seen with varying attenuation and thermal conductivity, but no visible changes were present with sound speed variations. The aberration study revealed an underlying trend in the spatial distribution of the phase shifts. \textit{Conclusion:} The results show that the efficacy of HIFU therapy in the kidney could be improved with aberration correction. \textit{Significance:} A method is proposed by which patient specific pre-treatment calculations could be used to overcome the aberration and therefore make ultrasound treatment possible.
\end{abstract}


\end{frontmatter}

\copyrighttext

\pagebreak









\section*{Introduction}

The incidence of kidney (renal) cancer has been growing at an annual rate of 2\% with the vast majority of the cases being renal cell carcinomas (RCC) \citep{luciani2000incidental, ljungberg2011epidemiology, globocan2012cancer}. In 2012 it was the 13$^{\mathrm{th}}$ most common cancer in the world \citep{globocan2012cancer} with approximately 338,000 new cases diagnosed (214,000 in men and 124,000 in women), representing 2.4\% of all cancers. In the same year approximately 143,000 people died due to the disease. The five-year survival rate of kidney cancer has been around 74\% in recent years, but patients with advanced RCC have five-year survival rates of only 11-12\% \citep{nci2016kidney}. Early diagnosis as well as safe and effective therapy methods are therefore crucial for improving patient outcomes. 

Improvements in diagnostic imaging modalities, such as ultrasound, magnetic resonance imaging (MRI) and computed tomography (CT), have benefited the early detection of kidney cancer, but effective treatment of the disease still remains a challenge. Typically kidney cancer is treated surgically, which is currently the only curative option available \citep{van2011prospective}, but it can lead to complications in as many as 19\% of cases \citep{gill2003comparative}. Alternative, minimally invasive therapies such as cryotherapy \citep{gill2005renal} and radiofrequency ablation \citep{gervais2005radiofrequency} reduce the risk of complications and often result in shorter hospital stays. However, neither of these methods is completely non-invasive and therefore still present a risk of infection, seeding metastases and other complications.

High-intensity focused ultrasound (HIFU) is a non-invasive therapy method which does not require puncturing the skin and typically has minimal or no side-effects. In HIFU therapy, focused ultrasound beams are used to create a rapid temperature rise at the focal point, which results in irreversible tissue damage due to coagulative thermal necrosis \citep{wall1999thermal, wright2002denaturation}. HIFU therapy can be used clinically to treat cancerous tissue in kidney, but the oncological outcomes have been variable \citep{wu2003preliminary, illing2005safety, marberger2005extracorporeal, ritchie2010extracorporeal}.

\citet{wu2003preliminary} demonstrated the feasibility of HIFU ablation of renal malignancies, all but one being RCC. A total of 13 patients were treated, of which 10 had partial ablation and three had complete tumour ablation. \citet{illing2005safety} also tested the safety and feasibility of HIFU renal ablation in eight patients. Four of the treated patients had surgical resection of the kidney after the treatment, of which only one showed features of ablation. In addition, six patients had a MRI assessment of the response and ablation was demonstrated in four. \citet{marberger2005extracorporeal} presented a clinical phase II trial results of extracorporeal ablation of renal tumours with 16 treated patients. They found acute tissue necrosis \citep{susani1992morphology} in nine tumours exposed to the highest dose of ultrasound, but this only covered 15-35\% of the targeted area. \citet{ritchie2010extracorporeal} showed in a study of 15 patients that only three had more than half the tumour ablated and eight had no detectable signs of ablation.

The variable degree of efficacy in HIFU ablation of the kidney could be due to two reasons: limitations in the therapeutic HIFU system and the physical factors related to the human body. With respect to the therapy system, the diameter of the HIFU transducer has to be relatively large, typically above 10 cm in diameter, which allows the pre-focal ultrasound beam energy to be spread over a wider area. This reduces the pre-focal heating and the possible effect of shielding, particularly from the rib cage. In addition, large diameter transducers allow for greater focal lengths which are up to 15 cm, thus allowing the treatment of deep-lying organs such as the kidney. Due to the location of the kidney, the ultrasound frequency also needs to be low enough that the attenuation from intervening tissue layers does not remove much energy. Therefore, extracorporeal HIFU systems typically operate in the frequency range of 0.5-1.5 MHz to maximise the ultrasound penetration depth with high enough intensity \citep{ter2007high}.

In addition to the requirements for the HIFU system, the structure and acoustic properties of tissues in front of the transducer affect the efficacy of HIFU therapy. Due to the deep location of the kidney, several tissue layers, including skin, fat, muscle and soft tissue, lie in front the kidney. These layers will reduce the intensity of the ultrasound field due to attenuation. The effect of attenuation might be particularly significant in the nonlinear case in which higher harmonic frequencies generated during HIFU therapy are more strongly attenuated. In addition to attenuation, the defocusing of ultrasound due to refraction and reflections at tissue interfaces might result in significant loss of HIFU energy in the target location. Kidneys are also highly perfused organs, which causes heat dissipation, and thus, reduced heating efficacy. Therefore, all the factors discussed above should be considered in order to achieve successful thermal ablation in the kidney.

The aim of this research was to investigate how the combined effect of attenuation, reflection and refraction of different tissue layers in front of the kidney affect the intensity and shape of the ultrasound field. This was done by performing nonlinear HIFU therapy simulations in segmented three-dimensional CT datasets of three different patients. A preliminary study of acoustic simulations in a single patient has been published in \citep{suomi2016nonlinear}. After the acoustic simulations, the heating efficacy of HIFU therapy in the kidney was determined with thermal simulations. The acoustic and thermal parameters as well as the perfusion of the kidney were varied within their physiological limits in order to examine their effect on heating. In addition, an aberration study to examine the effect of tissue layers on phase shifts at the transducer face was conducted. These results provide a quantitative analysis of the factors affecting the overall efficacy of HIFU therapy of the kidney.

\section*{Computational model}

\subsection*{Parallelised nonlinear ultrasound simulation model}

The acoustic simulations were performed using the open-source k-Wave Toolbox \citep{treeby2012modeling}. This solves a set of coupled first-order partial difference equations based on the conservation laws and a phenomenological loss term that accounts for acoustic absorption with a frequency power law of the form $\alpha = \alpha_0 \omega^y$ \citep{treeby2010modeling}. The governing equations can be written as:
\begin{align}
\frac{\partial \mathbf{u}}{\partial t} &= -\frac{1}{\rho_0} \nabla p \\
\frac{\partial \rho}{\partial t} &= -\left( 2\rho + \rho_0 \right) \nabla \cdot \mathbf{u} - \mathbf{u} \cdot \nabla \rho_0 \\
p &= c_0^2 \left( \rho + \mathbf{d} \cdot \nabla \rho_0 + \frac{B}{2A}\frac{\rho^2 }{\rho_0} - L\rho \right)
\end{align}
where $p$ is the acoustic pressure, $\mathbf{u}$ and $\mathbf{d}$ are the acoustic particle velocity and displacement, $\rho$ and $\rho_0$ are the acoustic and background density, $c_0$ is the isentropic sound speed, $B/A$ is the nonlinearity parameter, and $L$ is a loss operator accounting for acoustic absorption and dispersion that follows a frequency power law \citep{treeby2010modeling}:
\begin{align}
L =& -2\alpha_0 c_0^{y-1} \frac{\partial}{\partial t} \left( -\nabla^2 \right)^{\tfrac{y}{2}-1} \nonumber
\\ & + 2 \alpha_0 c_0^y \tan \left( \pi y /2 \right)\left( -\nabla^2 \right)^{\tfrac{y + 1}{2} - 1}
\end{align}
These expressions are equivalent to a generalised version of the Westervelt equation that accounts for second-order acoustic nonlinearity, power law acoustic absorption, and a heterogeneous distribution of material properties (sound speed, density, nonlinearity and absorption coefficient).

The governing equations were solved using a $k$-space pseudospectral meth\-od, where spatial gradients are calculated using the Fourier collocation spectral method, and time integration is performed using an explicit dispersion-corrected finite-difference scheme \citep{tabei2002k}. The model was implemented in C++ and optimised for distributed computing environments using the standard message passing interface (MPI) \citep{jaros2015full}. The 3-D domain was distributed across multiple cores using 1-D slab decomposition, and the MPI version of the FFTW library was used to perform the requisite  Fourier transforms \citep{frigo1998fftw}.

\subsection*{Thermal simulation model}

The thermal simulation model was constructed of a three-dimensional heat equation which took into account the nonlinear heating rate of the ultrasound field as well as the perfusion in the kidney. The coupled heat equation can be expressed as:
\begin{align}
\label{eq:heat_equation}
\rho_{\mathrm{k}} C_{\mathrm{k}} \frac{\partial T}{\partial t} = k_{\mathrm{k}} \nabla^{2}T - w_{\mathrm{k}} C_{\mathrm{b}}(T-T_{0}) + H
\end{align}
where $T$ is the three-dimensional ($x$, $y$, $z$) temperature field, $T_{0}$ the initial condition (here $T_{0}$ = 37~$^{\circ}$C); $H$ is the heating rate; $\rho$, $C$, $k$ and $w$ are the density, specific heat capacity, thermal conductivity and perfusion with the subscripts `k' and `b' referring to kidney and blood, respectively. Because the simulated ultrasound fields were nonlinear, the heating rate was calculated using the harmonic components of the acoustic field according to the equation:

\begin{align}
\label{eq:heat_flux}
H = \frac{1}{c_{\mathrm{k}} \rho_{\mathrm{k}}}\sum_{n=1}^{N} \alpha_{\mathrm{k}}(nf_{0}) |P_{n}|^{2}
\end{align}
where $c$ is the sound speed, $\alpha$ is the frequency dependent attenuation, $f_{0}$ is the sonication centre frequency, $P_{n}$ is the pressure of the harmonic component $n$ and $N$ is the number of harmonics (here $N$ = 4). The pressure values of each harmonic component were obtained using the discrete Fourier transform (DFT) of the time-domain ultrasound waveforms at each spatial location. The heat equation was solved using the alternating direction implicit (ADI) method \citep{ames1977numerical}.

%

\section*{Simulations}

\subsection*{Therapeutic high-intensity focused ultrasound simulations}

\begin{figure}[b!]
    \centering
        \subfigure[]
    {
        \includegraphics[height=4cm]{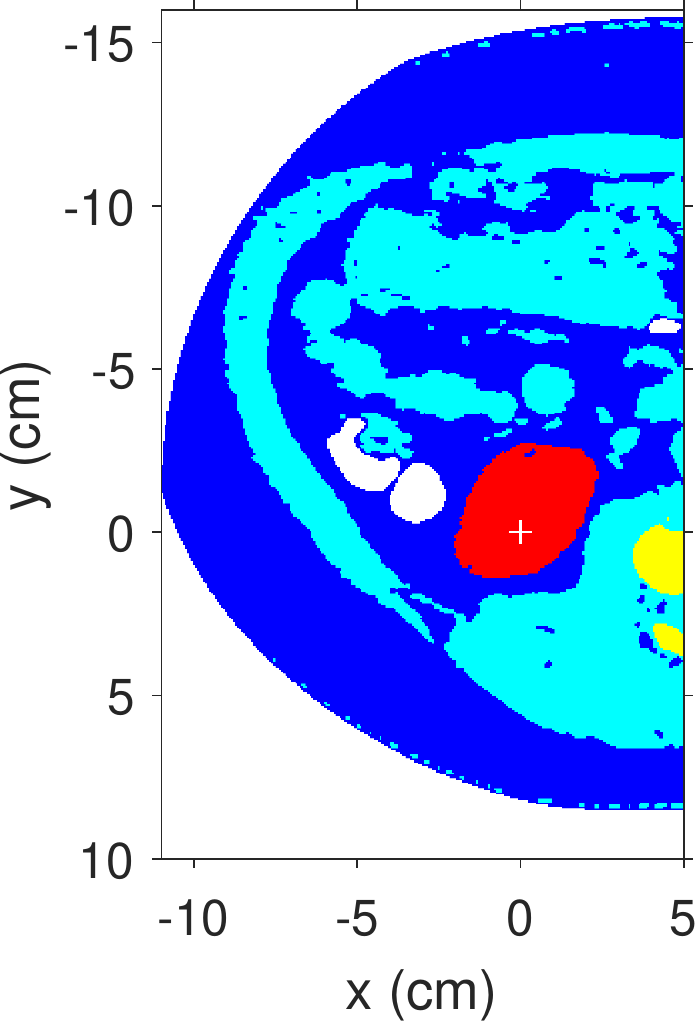}
    }
    \subfigure[]
    {
        \includegraphics[height=4cm]{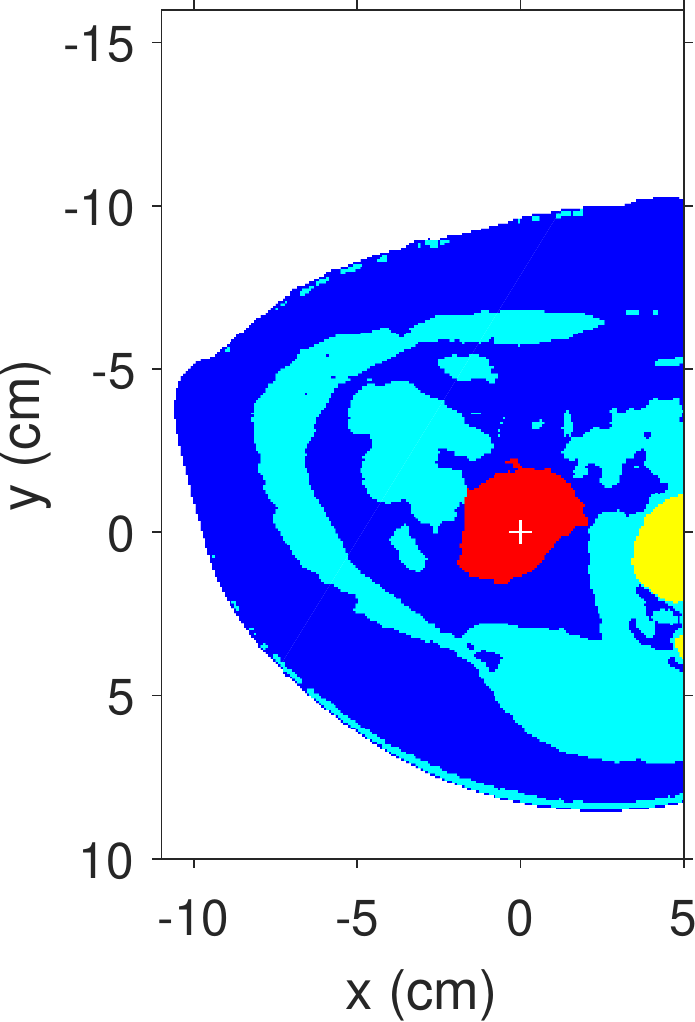}
    }
    \subfigure[]
    {
        \includegraphics[height=4cm]{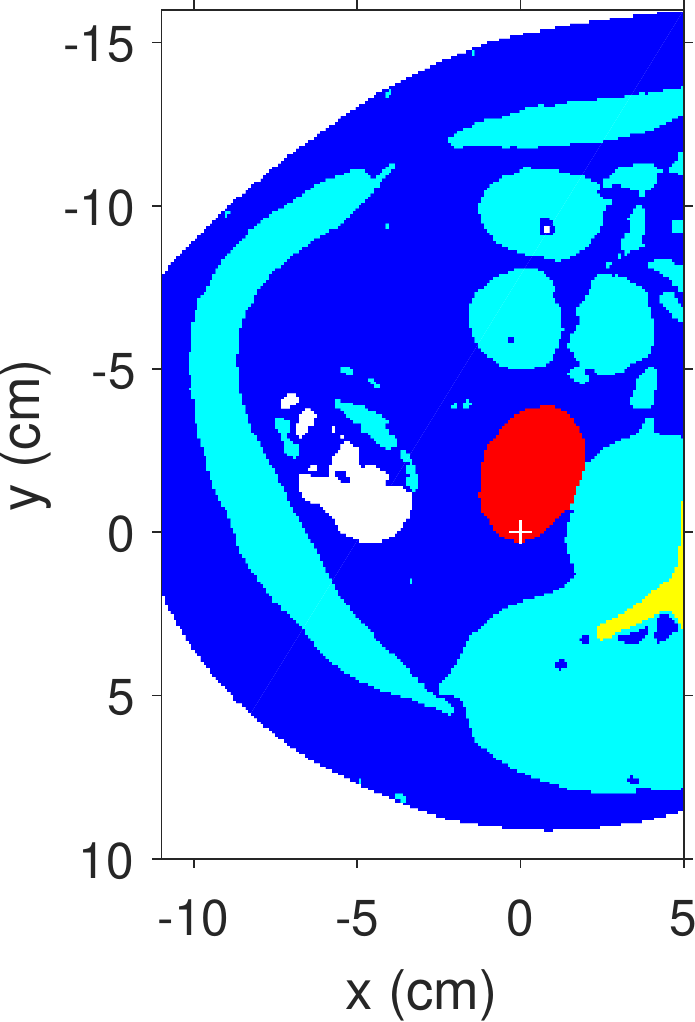}
    }
    \\
    \subfigure[]
    {
        \includegraphics[height=4cm]{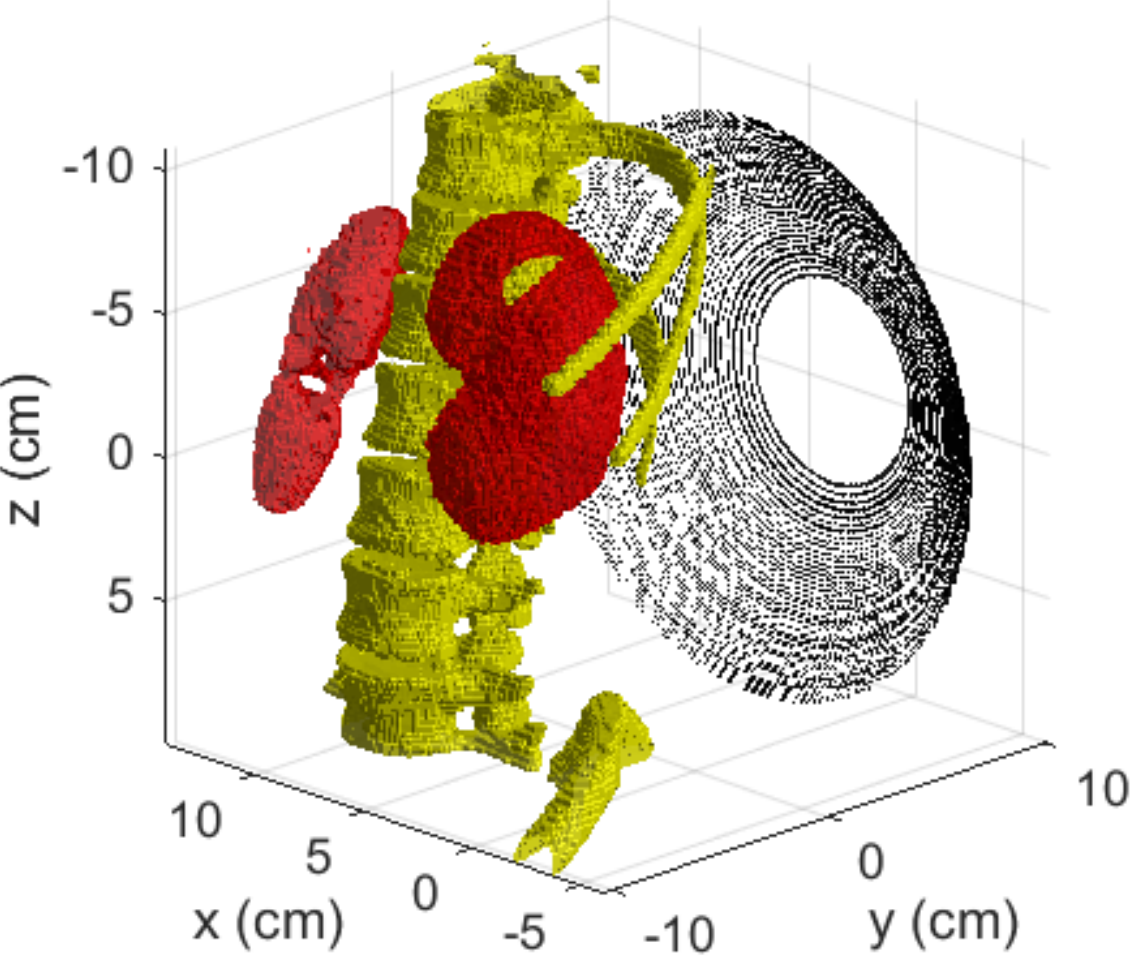}
    }
    \subfigure[]
    {
        \includegraphics[height=4cm]{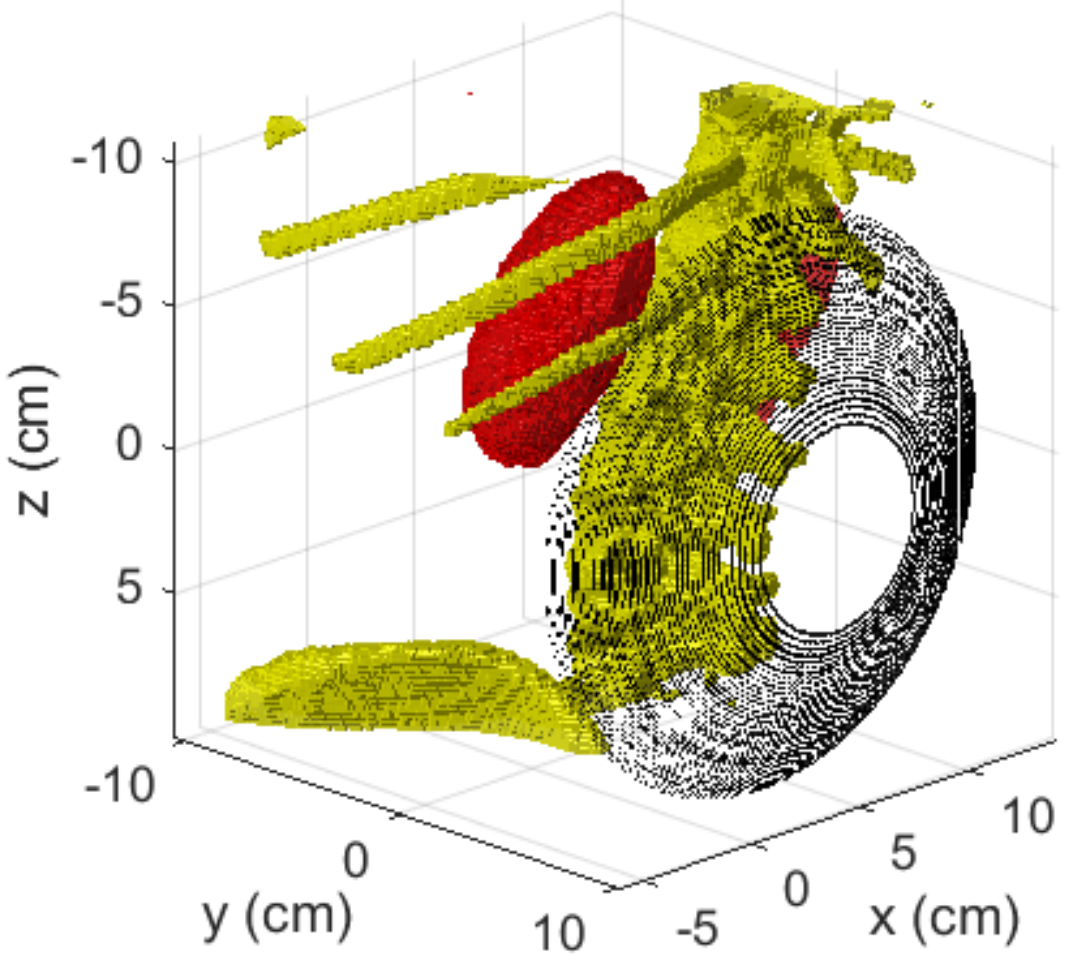}
    }
    \caption{(a)-(c) Axial segmented computed tomography (CT) slices of patients 1-3, respectively. The different colours in the segmented CT data correspond to medium type: white - water, yellow - bone, cyan - soft tissue, blue - fat and red - kidney. The ultrasound focal point target location is marked with a white cross. (d)-(e) 3-D visualisation of the CT scan in patient 1 showing the simulation geometry (without soft tissue, fat and water). Similar geometries targeting the lower part of left kidney were used for patients 2 and 3.}
    \label{fig:ct_3D}
\end{figure}

The simulation geometries were derived using CT datasets of three different patients (see Figure \ref{fig:ct_3D}). The patients were of different size with peri-nephric fat, subcutaneous fat and soft tissue in front of the kidney ranging between 0.4-1.6 cm, 1.8-2.6 cm and 3.0-5.0 cm, respectively. Thresholds were used to automatically segment the datasets into bone, fat and other soft tissue after which the kidneys were segmented manually. The medium outside the patients was segmented as water. Typical values reported by \citet{mast2000empirical} for sound speed, attenuation, density and B/A were used for each tissue type (see Table \ref{tab:acoustic_parameters}).

\begin{table}[t!]
  \centering
  \caption{Acoustic simulation parameters for different tissue types}
    \begin{tabular}{lcccc}
    \hline
          & Density   		& Sound speed   & Attenuation 				& B/A 	\\
          & (kg/m$^{3}$) 	& (m/s) 		& (dB/MHz$^{1.1}$/cm)		& 		\\
    \hline
    Water & 1000  			& 1520  		& 0.00217 					& 5.2 	\\
    Bone  & 1908  			& 4080  		& 20.00    					& 7.4 	\\
    Soft tissue & 1055  	& 1575  		& 0.60   					& 7.0 	\\
    Fat   & 950   			& 1478  		& 0.48  					& 10.0 	\\
    Kidney& 1050  			& 1560 $\pm$ 10	& 1.00 $\pm$ 0.24   		& 7.4 	\\
    \hline
    \end{tabular}
  \label{tab:acoustic_parameters}
\end{table}

The HIFU transducer was modelled on a clinical system (Model JC200, Haifu, Chongqing, China) with an annular transmitting surface of outer diameter 20 cm, inner hole diameter 6 cm, operating frequency 0.95 MHz and focal length 14.5 cm \citep{ritchie2013attenuation}. The transducer was positioned so that the geometric focal point of the transducer (the white cross in Figures \ref{fig:ct_3D}(a)-(c)) was located in the bottom part of the left kidney. This was done in order to avoid the ribs which would otherwise cause significant pressure losses during sonication due to strong reflection.

A reference simulation was carried out in water and two additional simulations for each patient: (i) with all tissue properties varying; and (ii) constant sound speed in all tissues to remove refraction, but all other properties varying. Four additional simulations for patient 1 were conducted by changing the attenuation and sound speed of the kidney by $\pm$0.24 dB/cm and $\pm$10 m/s, respectively, which correspond to $\pm$2 standard deviations (SD) based on 30 kidney samples in humans \citep{turnbull1989ultrasonic}.

Before performing the actual simulations, several convergence simulations were conducted in order to find the optimal grid size and temporal resolution. The computational grid consisted of 1200 $\times$ 1200 $\times$ 1200 grid points (i.e., 22.2 cm $\times$ 22.2 cm $\times$ 22.2 cm) giving a spatial resolution of 185 $\mu$m which supported nonlinear harmonic frequencies up to 4 MHz. Perfectly matched layers (PML) were used on the edges of the grid. The simulations were run as continuous wave and the time duration was set to 260 $\mu$s with a temporal resolution of 8.15 ns giving a total of 31876 time steps per simulation. The simulations were run using 400 computing cores for approximately 50 hours per simulation and requiring 200 GB of memory. The simulations were conducted using the computing facilities provided by advanced research computing (ARC) at the University of Oxford \citep{richards2016arc}. For data analysis, the time-domain waveforms and the peak pressures were saved in a three-dimensional grid around the focal point in each case. In addition, axial, sagittal and coronal slices of the ultrasound field over the whole spatial domain were saved.

\subsection*{Thermal simulations}

Thermal simulations were conducted in Matlab (R2015b, MathWorks Inc, Natick, MA, USA) using the nonlinear ultrasound fields from the acoustic simulations. The simulations were run for each patient in the kidney with the parameters presented in Table \ref{tab:thermal_parameters} \citep{hasgall2015database, roberts1995renal}. The thermal conductivity, perfusion rate of the kidney medulla and cortex in patient 1 were also changed by $\pm$0.04 W/m/K, $\pm$8.3 kg/m$^{3}$/s and $\pm$18.3 kg/m$^{3}$/s, respectively, which correspond to a $\pm$2 SD change \citep{hasgall2015database, roberts1995renal}. This was done in addition to the acoustic simulations in patient 1 which already included changing the attenuation and sound speed of the kidney. The thermal simulations were conducted in a 2 cm $\times$ 2 cm $\times$ 2 cm spatial domain around the target focal point (i.e., in the kidney) with a fixed temperature (Dirichlet) boundary condition of 37~$^{\circ}$C on the edges. Each sonication was conducted for 2 seconds which was followed by a 10-second cooling period. The evolution of the maximum temperature was recorded throughout the whole duration of the simulation. Furthermore, the temperature and cumulative equivalent minutes at 43~$^{\circ}$C (CEM$_{43 ^{\circ}\mathrm{C}}$), which is a measure of thermal dose \citep{sapareto1984thermal}, over the whole domain at the end of the cooling period were saved. 

\begin{table}[t!]
  \centering
  \caption{Thermal simulation parameters for kidney and blood}
    \begin{tabular}{lccc}
    \hline
          			& Thermal			& Specific		& Perfusion 		\\
          			& conductivity		& heat capacity	& rate				\\
          			& (W/m/K) 			& (J/kg/K) 		& (kg/m$^{3}$/s)	\\
    \hline
    Kidney (medulla)& 0.53 $\pm$ 0.04  	& 3763  		& 9.2 $\pm$ 8.3		\\
    Kidney (cortex)	& 0.53 $\pm$ 0.04  	& 3763  		& 46.3 $\pm$ 18.3	\\
    Blood 			& N/A      			& 3617  		& N/A 				\\
    \hline
    \end{tabular}
  \label{tab:thermal_parameters}
\end{table}

%

\section*{Results}

\subsection*{Ultrasound waveforms in time and frequency domain}

\begin{figure}[b!]
    \centering
    \subfigure[]
    {
        \includegraphics[width=0.46\columnwidth]{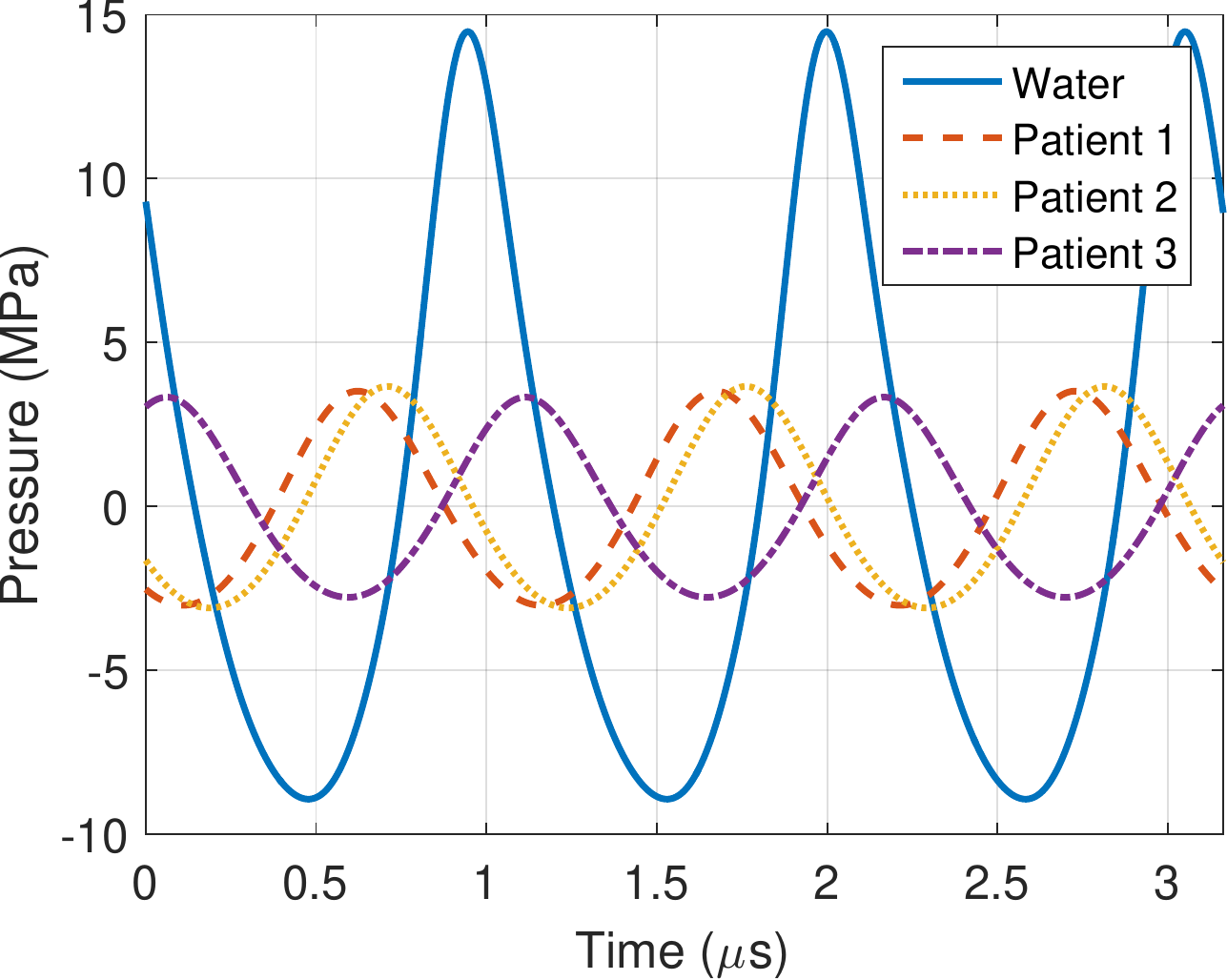}
    }
    \subfigure[]
    {
        \includegraphics[width=0.46\columnwidth]{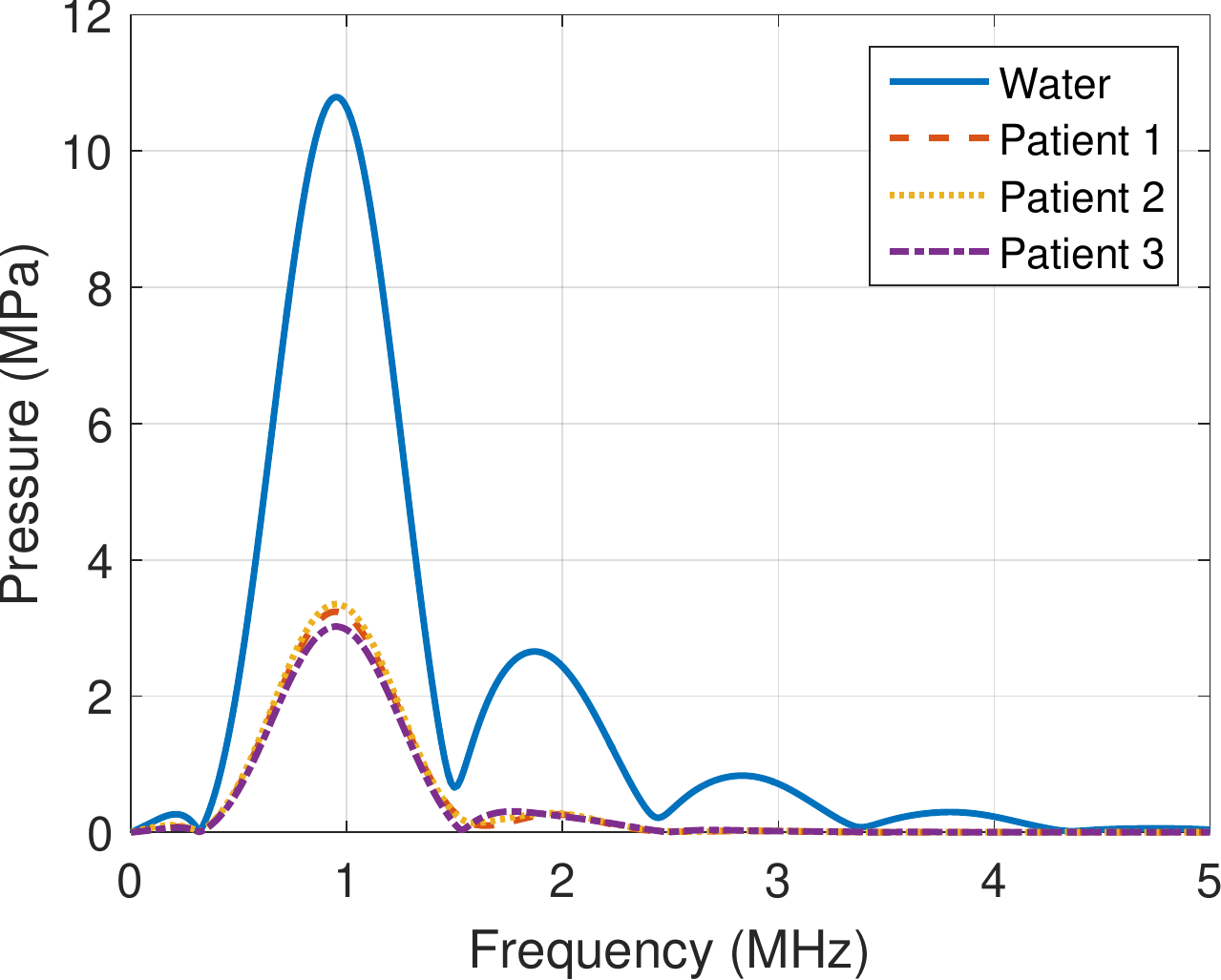}
    }
    \caption{(a) Time domain waveforms at the peak pressure location in water and kidney in patients 1-3 with refraction effects. (b) Windowed (Hann) frequency spectrum of the same waveforms.}
    \label{fig:waveform_time_fft}
\end{figure}

Figure \ref{fig:waveform_time_fft}(a) shows time domain waveforms at the location of the global peak pressure in water and kidney for the three different patients including refraction.
The individual peak pressures and spatial peak-temporal average intensity $I_{\mathrm{SPTA}}$ values for each case are also listed in Table \ref{tab:acoustic_results}. In water the peak pressure was 14.49 MPa with a corresponding $I_{\mathrm{SPTA}}$ of 4116 W/cm$^{2}$. In all patients, the mean peak pressure was 3.50 MPa which corresponds to an approximately 76\% drop in pressure amplitude. The range was from 3.33 to 3.66 MPa, which suggests that there is not much difference in the outcome between the three different patients. Similarly, the $I_{\mathrm{SPTA}}$ in patients dropped to an average value of 318 W/cm$^{2}$, which corresponds to a 92\% or an 11.1 dB drop. The range varied from 283 to 346 W/cm$^{2}$, which shows an 11\% variation around the mean, suggesting that heating should not vary dramatically across patients.

\begin{table}[t!]
  	\centering
  	\caption{Acoustic simulation results}
  	\begin{adjustbox}{max width=\columnwidth}
  	\begin{threeparttable}
    \begin{tabular}{lccccccc}
    \hline
     			& Peak		& $I_{\mathrm{SPTA}}$		& $I_{\mathrm{SPTA}}$	& \multicolumn{2}{c}{Focal shift}& \multicolumn{2}{c}{Parent focal size} 	\\
     			& pressure	& 			& reduction& Axial	& Radial & Length& Width 	\\
     			& (MPa)		& (W/cm$^2$)& (dB) 	& \multicolumn{2}{c}{(mm)} 	& \multicolumn{2}{c}{(mm)}		\\
    \hline
    Water 		& 14.49   	& 4116 	& 0.0	& 0.0	& 0.0		& 6.5	& 1.3		\\
    \hline
    Patient 1 	& 3.51  & 324 		& $-$11.0& 2.6	& 1.8		& 11.4	& 1.8		\\
    Patient 2 	& 3.66  & 346 		& $-$10.8& 0.6	& 1.2		& 9.5	& 1.4		\\
    Patient 3 	& 3.33  & 283		& $-$11.6& 3.1	& 1.1		& 7.3	& 1.2		\\
    \hline
    \textbf{Average} 	& \textbf{3.50} & \textbf{318}	& \textbf{$-$11.1}& \textbf{2.1}&	\textbf{1.4}	& \textbf{9.4}	& \textbf{1.5} \\
    \textbf{SD}	& \textbf{0.13}&	\textbf{26}& \textbf{0.4}&\textbf{1.1}&	\textbf{0.3}	&\textbf{1.7}&	\textbf{0.2}\\
    \hline
    Patient 1$^{*}$& 6.46	&	957	& $-$6.3	& 1.3	& 0.4	& 7.4	& 1.5	\\
    Patient 2$^{*}$& 6.25	&	887	& $-$6.7	& 1.3	& 0.3	& 7.4	& 1.4	\\
	Patient 3$^{*}$& 6.67	&	971	& $-$6.3	& 0.9	& 0.3	& 7.2	& 1.3	\\
	\hline
	\textbf{Average}& \textbf{6.46}	& \textbf{938}	& \textbf{$-$6.4}	& \textbf{1.2} 	& \textbf{0.3}	& \textbf{7.3}	& \textbf{1.4}	\\
	\textbf{SD}		& \textbf{0.17}	& \textbf{37}	& \textbf{0.2}	& \textbf{0.2}	& \textbf{0.1}	& \textbf{0.1}	& \textbf{0.1}	\\
	\hline
    \end{tabular}
    \begin{tablenotes}
  	\item[*] simulation without refraction effects
  	\end{tablenotes}
  	\end{threeparttable}
    \end{adjustbox}
  \label{tab:acoustic_results}
\end{table}

When no refraction effects were included in the simulations, the peak pressure and $I_{\mathrm{SPTA}}$ increased in all three patients, with an average value of 6.46 MPa (i.e., a 55\% drop in amplitude), which is approximately twice higher than that with refraction. The range was from 6.25 to 6.67 MPa indicating 3\% variation around the mean value suggesting that refraction is responsible for much of the 11\% variation seen in the refraction case. The average $I_{\mathrm{SPTA}}$ was 938 W/cm$^{2}$ which corresponds to a 77\% or a 6.4 dB drop in intensity. In this case, patient 2 had the lowest $I_{\mathrm{SPTA}}$ of 887 W/cm$^{2}$ with patients 1 and 3 having 957 W/cm$^{2}$ and 971 W/cm$^{2}$, respectively. The small range suggests that heating should be very similar in these three patients.

Figure \ref{fig:waveform_time_fft}(b) shows the frequency spectra of the same focal waveforms in water and the three patients with refraction. The harmonics have broad lobes due to windowing (three cycles). A peak at the centre frequency of 0.95 MHz is clearly visible in each case as are the nonlinearly generated harmonics. However, in the case of tissue, the nonlinear effects are much less pronounced when compared to water. In water the second harmonic is 25\% of the fundamental component compared to 7\% in tissue. All three patients show a low magnitude second harmonic while the third and fourth harmonics are barely visible. These data suggest that for this HIFU system nonlinearity does not play a major role in heating.

\subsection*{Ultrasound pressure fields}

\begin{figure}[b!]
    \centering
    \subfigure[]
    {
        \includegraphics[height=3cm]{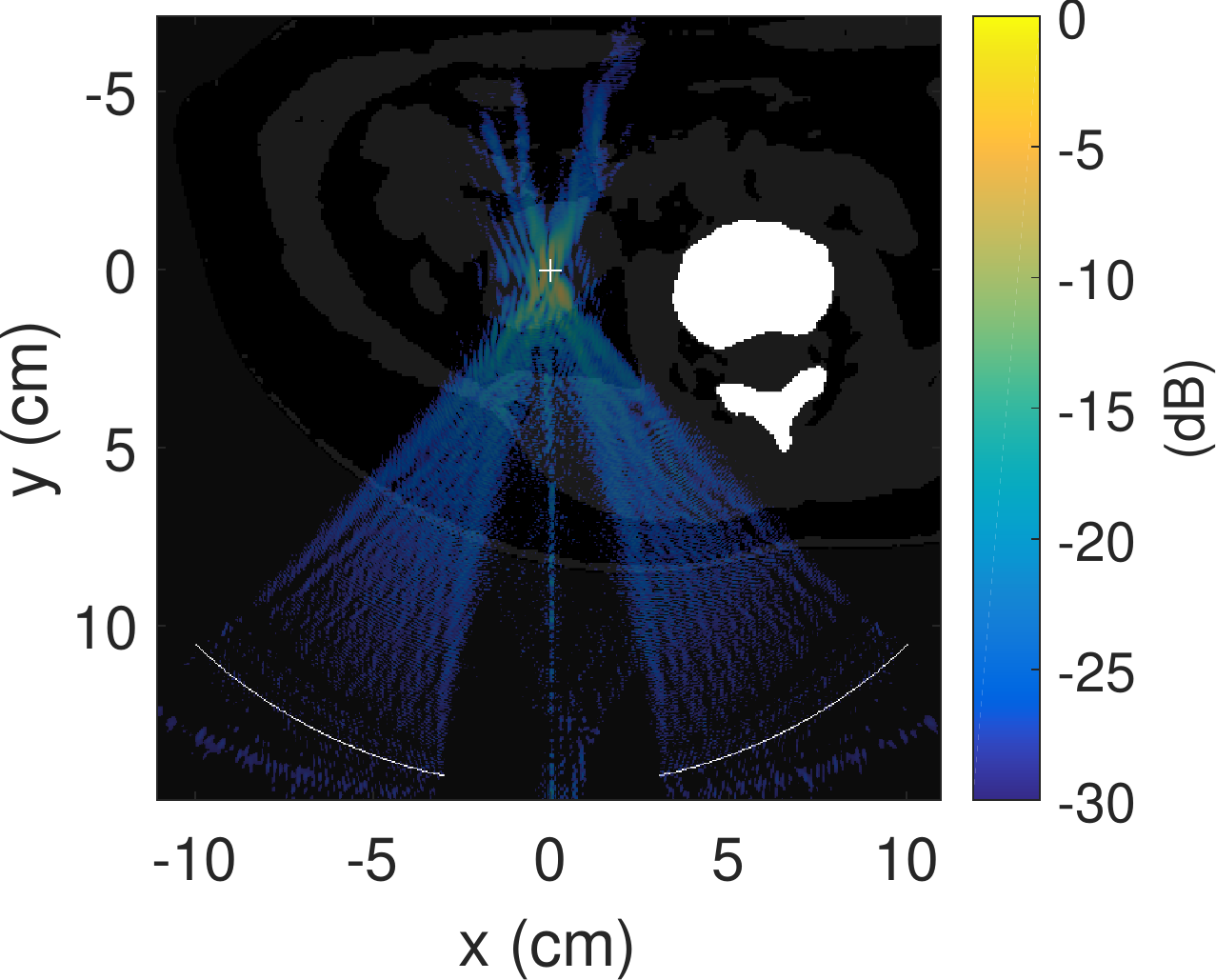}
    }
    \subfigure[]
    {
        \includegraphics[height=3cm]{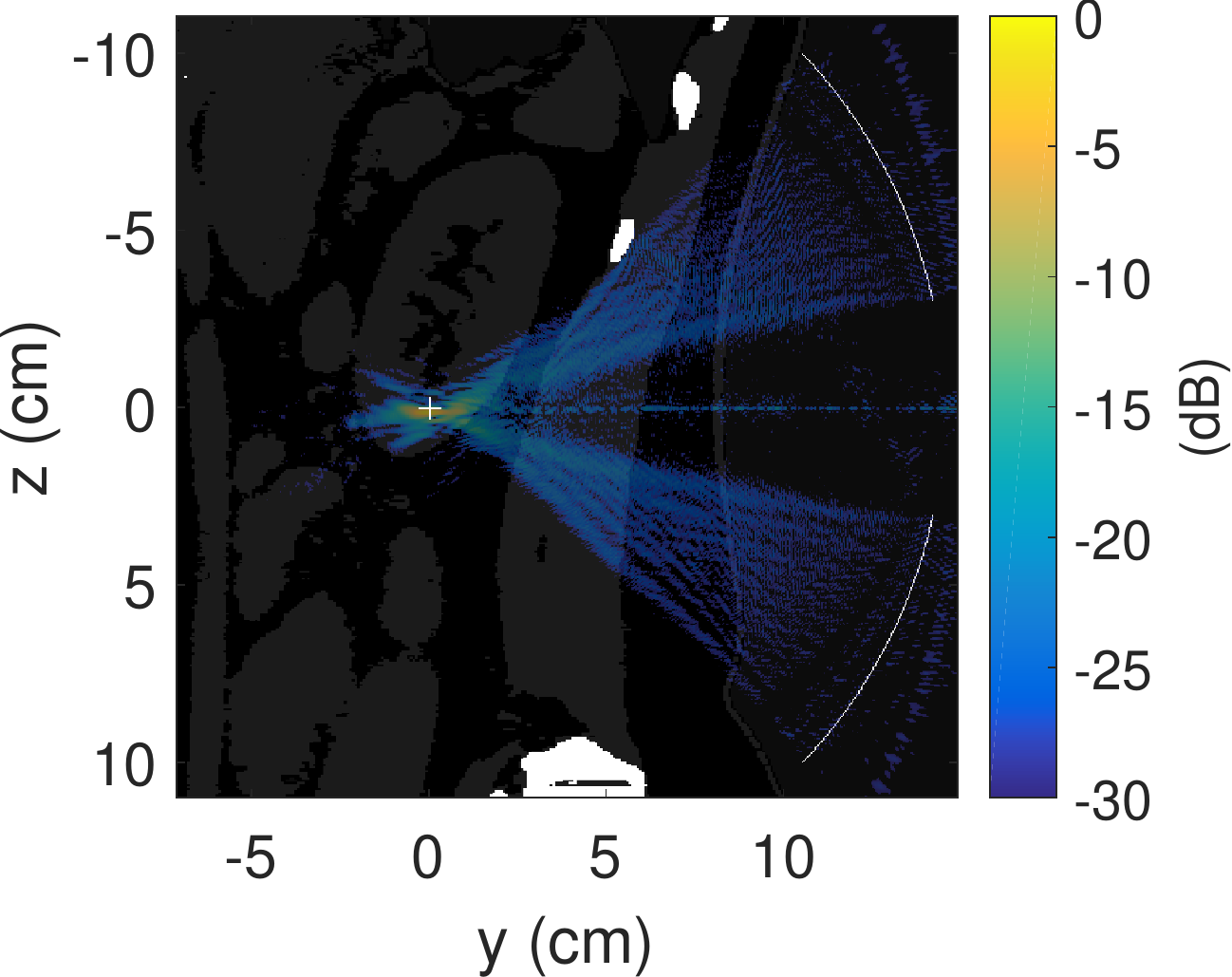}
    }
    \subfigure[]
    {
        \includegraphics[height=3cm]{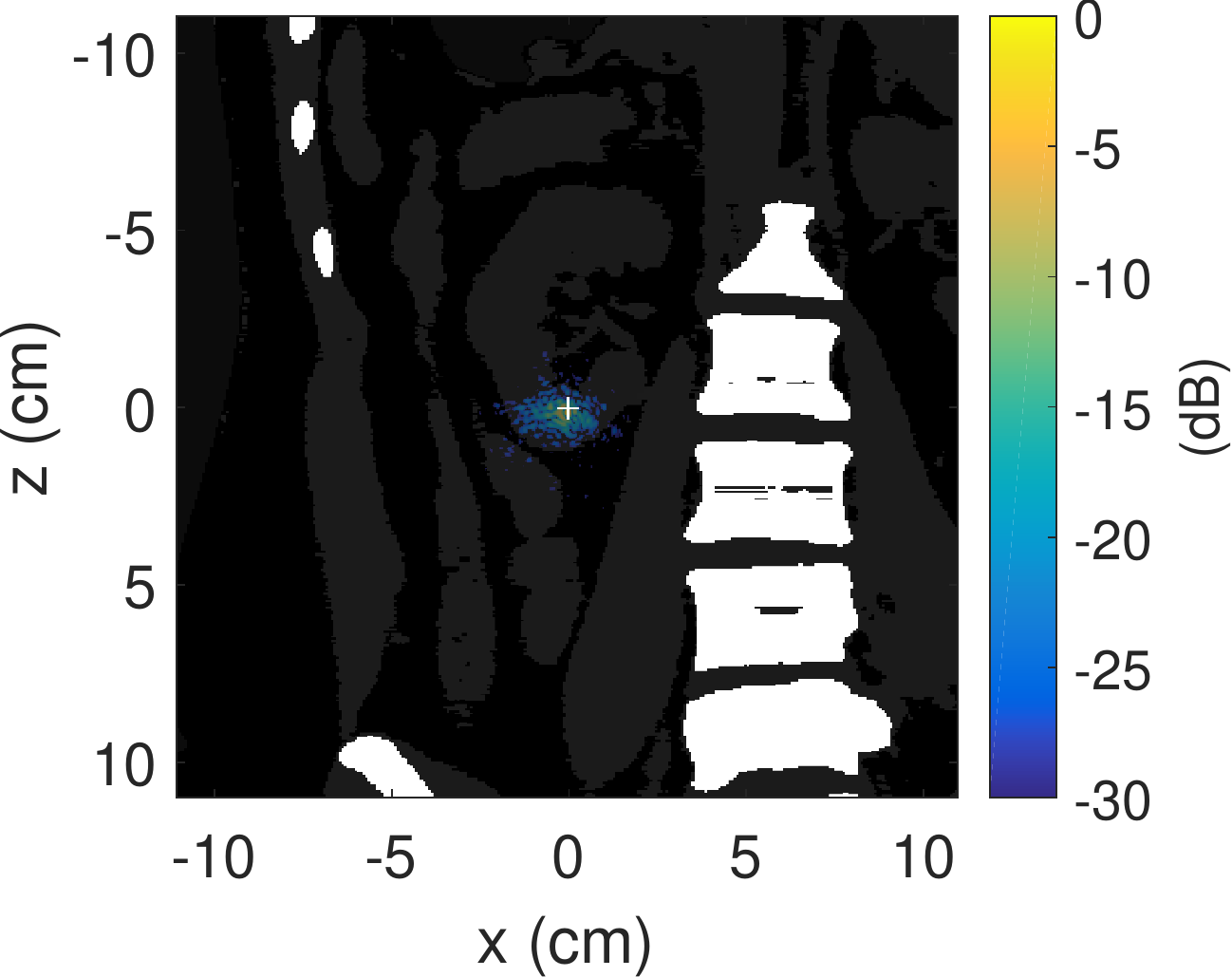}
    }
    \\
    \subfigure[]
    {
        \includegraphics[height=3cm]{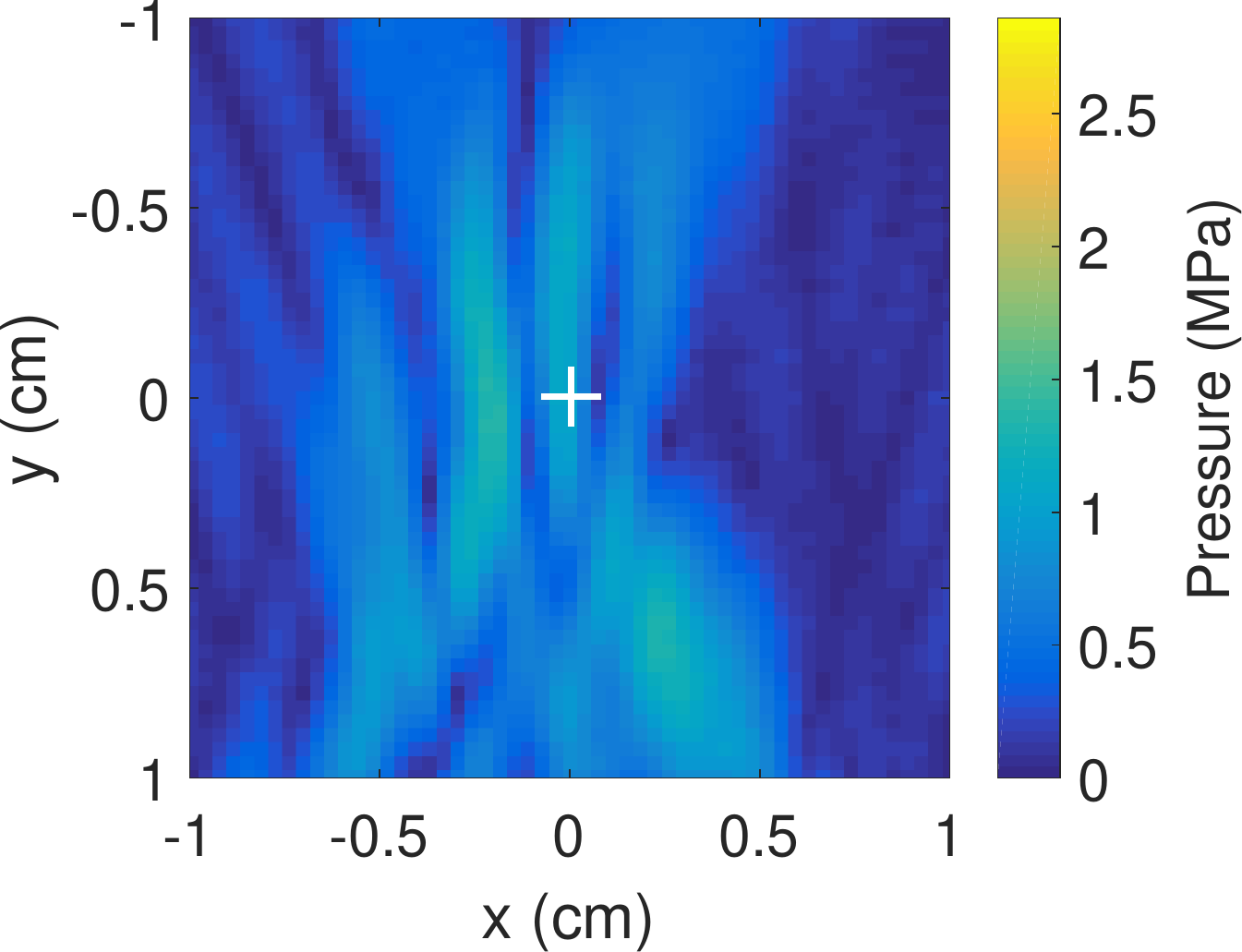}
    }
    \subfigure[]
    {
        \includegraphics[height=3cm]{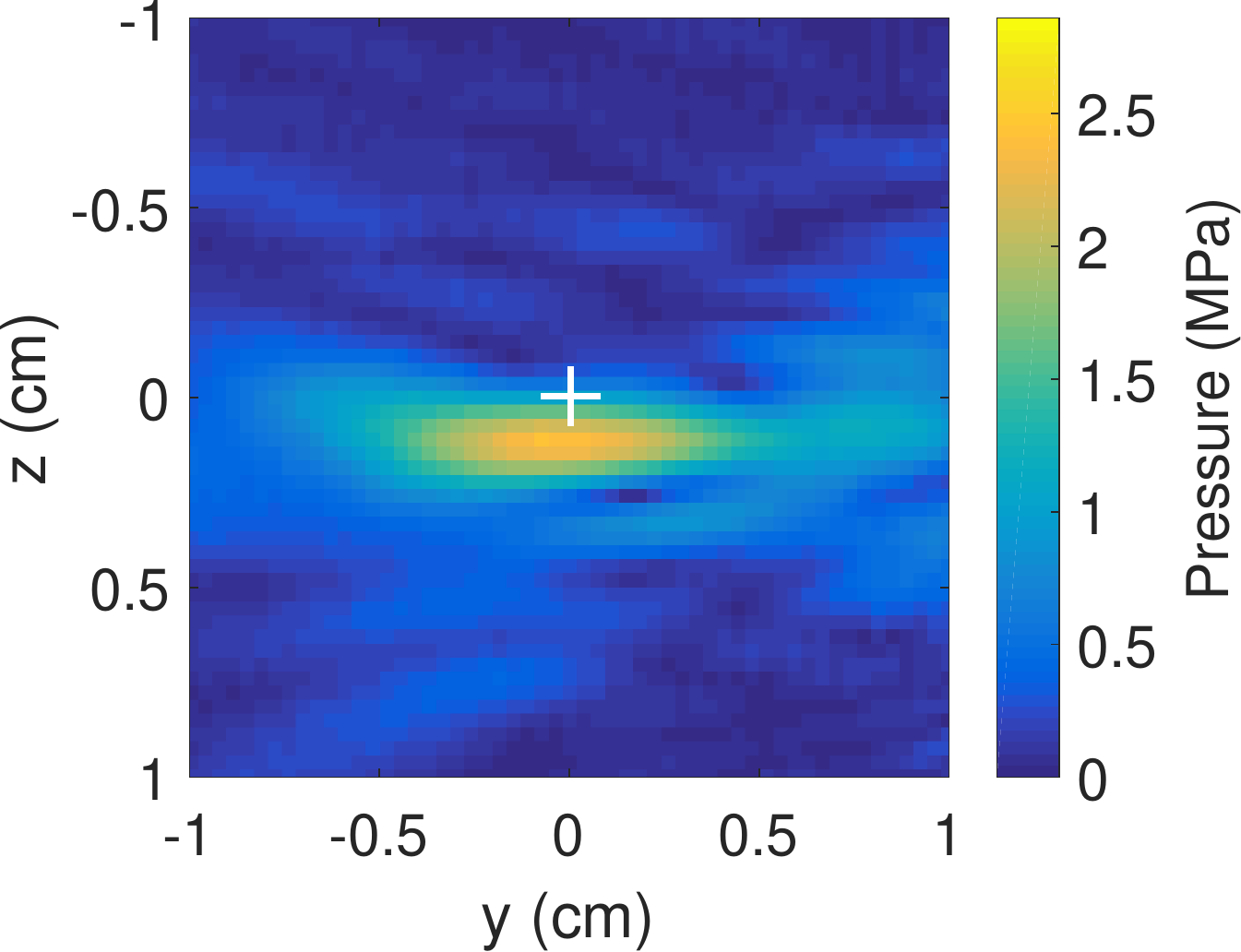}
    }
    \subfigure[]
    {
        \includegraphics[height=3cm]{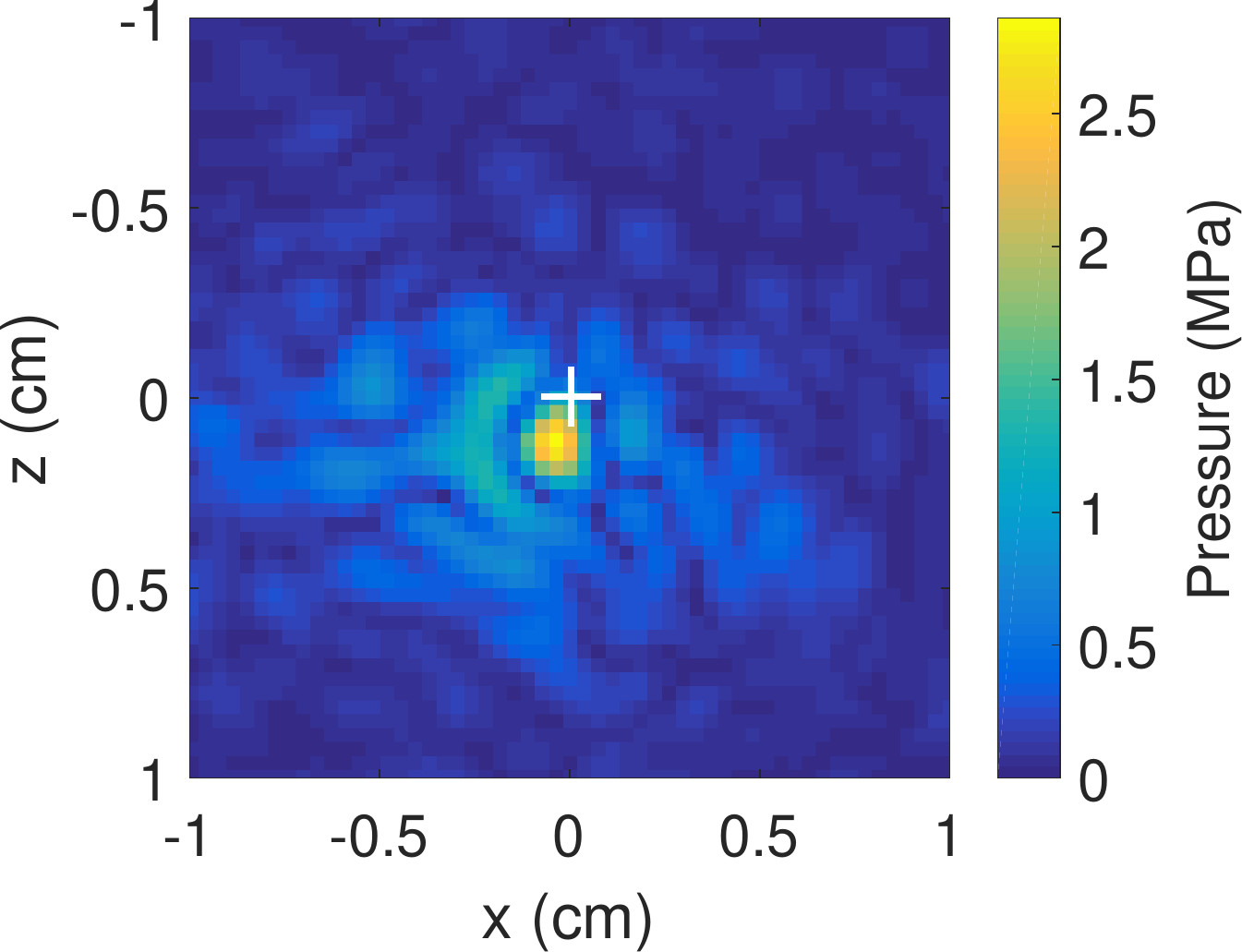}
    }
    \caption{(a) Axial, (b) sagittal and (c) coronal slices of the computed tomography (CT) scan showing the ultrasound pressure field in patient 2. The pressure field is displayed on a log-scale with a dynamic range of 30 dB. The ultrasound focal point target location is marked with a white cross. (d) Axial, (e) sagittal and (f) coronal slices of the ultrasound field in the focal area in the kidney on a linear pressure scale.}
    \label{fig:us_focal}
\end{figure}

Figures \ref{fig:us_focal}(a), (b) and (c) show the axial, sagittal and coronal slices of the ultrasound pressure field generated by the HIFU transducer in patient 2.
The pressure fields are displayed using a log-scale thresholded at $-$30 dB below the maximum pressure in each slice. The annular nature of the ultrasound source results in the shadow region in the centre of the beam. In the focal area, it can be seen that the region of high pressure does not form the archetypical ellipse shape, but is more diffuse instead. Furthermore, the areas of high pressure are offset from the focal point target location in all slices.

\begin{figure}[htbp!]
    \centering
    \subfigure[]
    {
        \includegraphics[height=5cm]{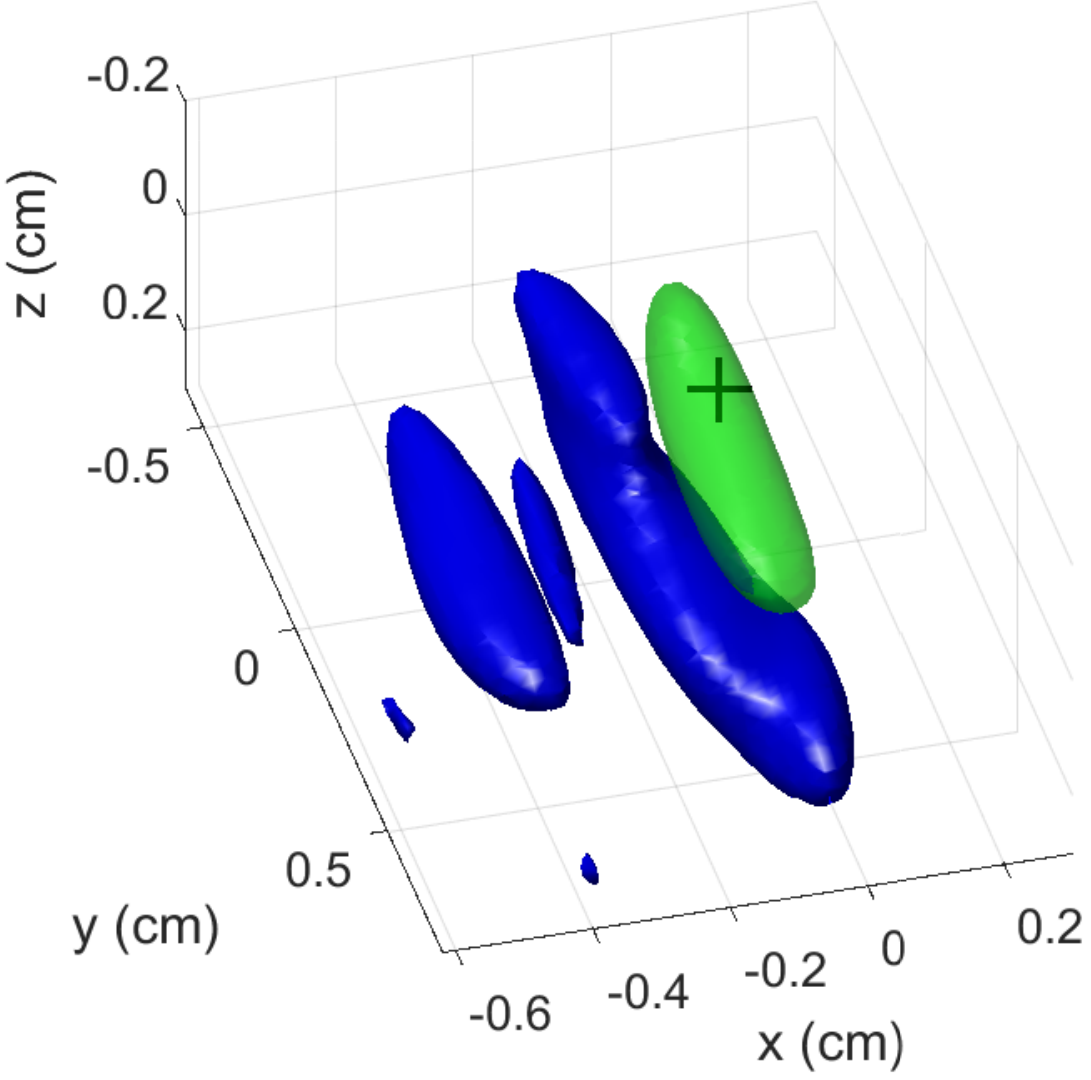}
    }
    \subfigure[]
    {
        \includegraphics[height=5cm]{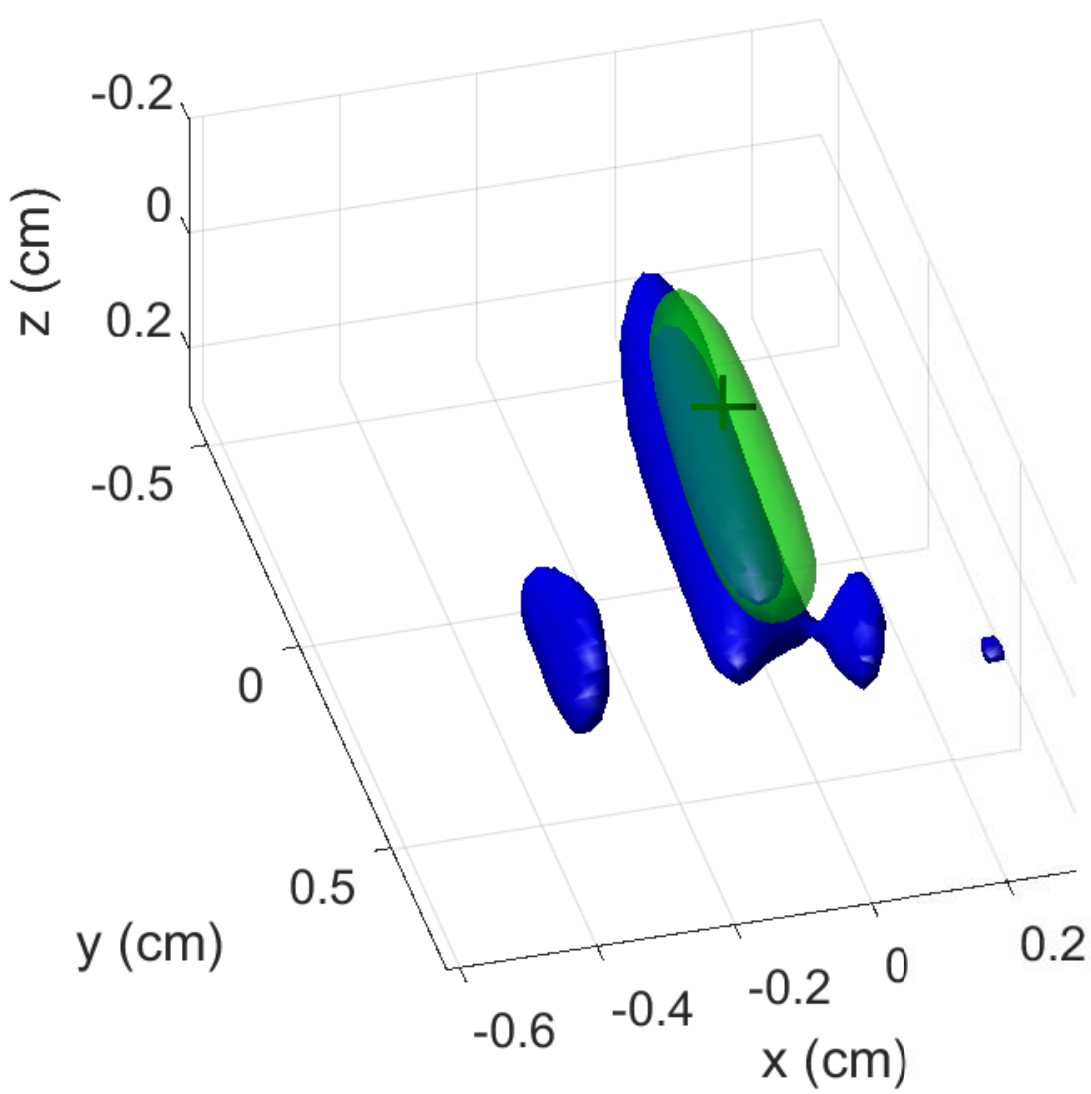}
    }
    \subfigure[]
    {
        \includegraphics[height=5cm]{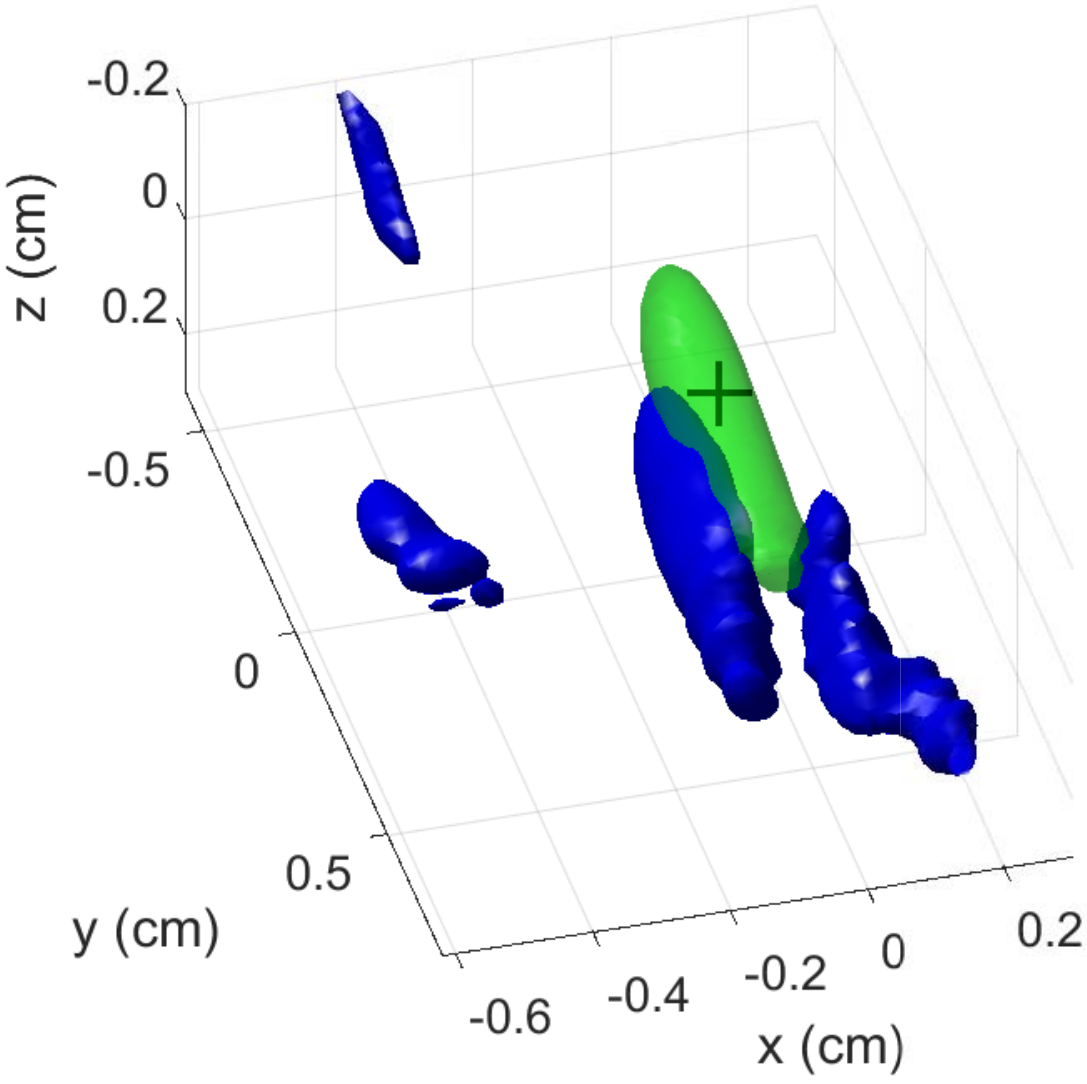}
    }
    \caption{(a)-(c) The $-$6 dB focal point volumes in the kidney for patients 1-3, respectively. The simulations with refraction are shown with blue isosurfaces while the simulations without refraction are shown with transparent green isosurfaces. The target focal point is marked with a black cross. The shifting and splitting of the focal point into one parent and several child focal volumes due to refraction can be seen in different patients.}
    \label{fig:focal_point_3D}
\end{figure}

Figures \ref{fig:us_focal}(d), (e) and (f) show close-ups of the axial, sagittal and coronal slices of the pressure field in the ultrasound focal area. It can be seen that the peak pressure does not occur at the target location (the white cross). This effect was observed in all three patients and the offsets are given in Table \ref{tab:acoustic_results}. On average the shifts were observed to be 2.1 mm in the axial and 1.4 mm in the radial directions. By examining the focal area in more detail in the coronal slice in Figure \ref{fig:us_focal}(f), it can be seen that in addition to the focal shifting, a region of high pressure has split into a number of less well-defined subvolumes. 

The splitting of the focal point is more clearly visualised in Figures \ref{fig:focal_point_3D}(a)-(c), which show the isosurfaces of the focal pressure regions thresholded at $-$6 dB in the three patients. The simulations with refraction are shown with blue isosurfaces while the simulations without refraction are green and transparent. The target focal point is marked with a black cross. For simulations with refraction, the largest $-$6 dB focal volume was identified as the parent focal volume and the others as child volumes. The parent focal volume lengths and widths are presented in Table \ref{tab:acoustic_results} for all three patients. In the case of patient 1 in Figure \ref{fig:focal_point_3D}(a), it can be seen that the focal region consists of five focal volumes with the largest (i.e., the parent) being approximately 11.4 mm in length and 1.8 mm in width. The corresponding values were observed to be 9.5 and 1.4 mm in patient 2 and 7.3 and 1.2 mm in patient 3. On average the parent focal size was 9.4 mm in length and 1.5 mm in width. Without refraction there was no splitting of the focal volume and the focus coincided with the target. The average values for length and width of the focal point without refraction were 7.3 mm and 1.4 mm which are 22\% and 7\% smaller, respectively. For comparison, the size of the $-$6 dB focal point in water was approximately 6.5 mm in length and 1.3 mm in width. This indicates that it is refraction that dominates the shifts and splitting of the focus.

The splitting of the focal region was quantified by comparing the size and pressure distribution in the child volumes to those of the parent volume in each patient. This was done as an indicator of the heating efficacy of the child volumes. For a given pressure bin (defined between 50\% and 100\% of the global maximum pressure with 10\% bin width) the cumulative volume of the child voxels in the bin was compared to the volume of all voxels above 50\% in the parent volume.

\begin{figure}[b!]
    \centering
    \includegraphics[height=5cm]{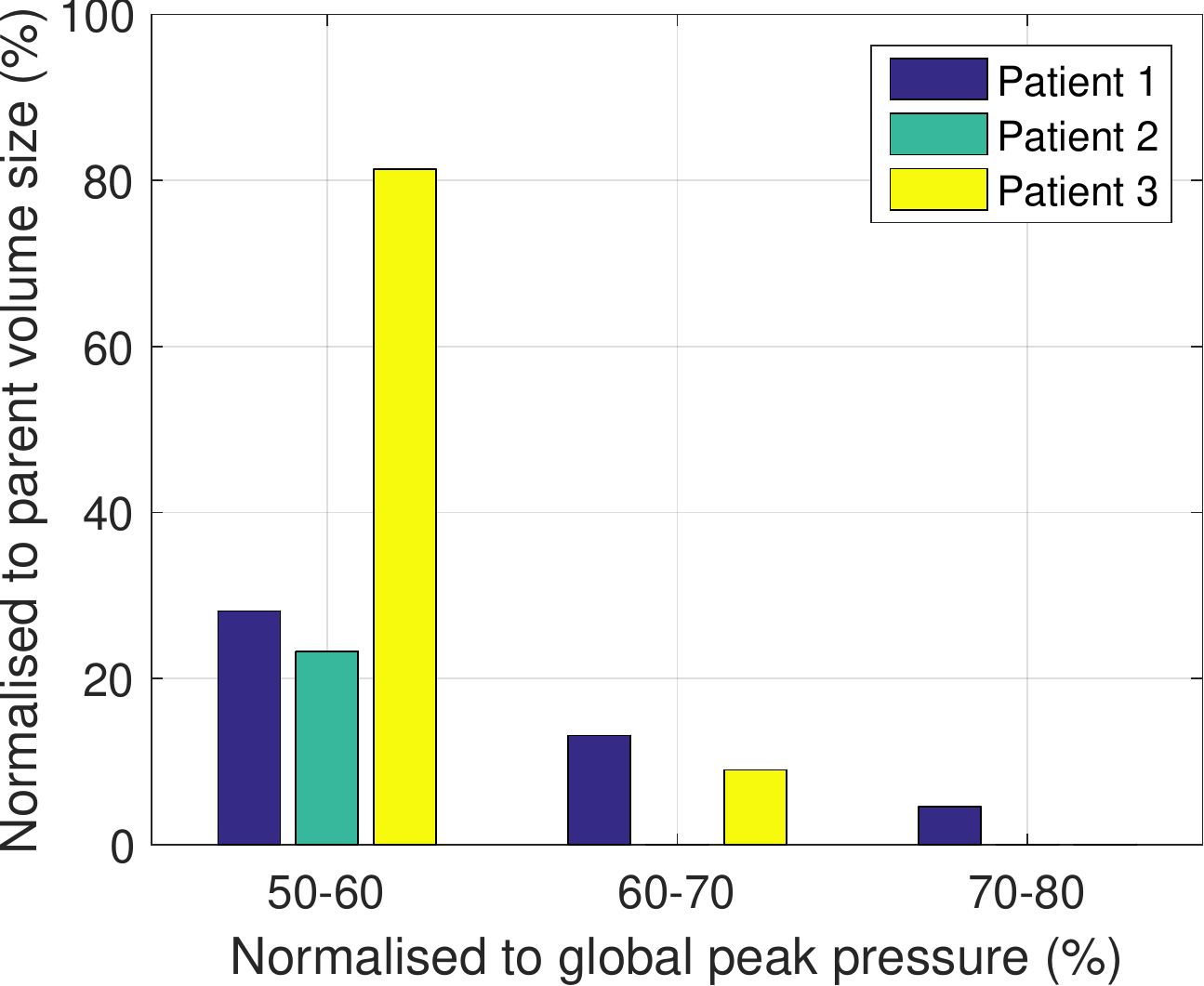}
    \caption{Histogram showing the pressure distribution in the child volumes with respect to the parent focal volume (i.e., the largest blue volumes) with bins varying from 50\% to 80\% of the global peak pressure in each case.}
    \label{fig:histogram_3D}
\end{figure}

Figure \ref{fig:histogram_3D} shows a histogram of the analysis for three different patients. In the 50-60\% pressure region, the cumulative size of child voxels was approximately 28\% of the parent focal point in patient 1. In patient 2 the same value was 23\% while in patient 3 a considerably higher value of 81\% was observed. This is also apparent from Figure \ref{fig:focal_point_3D}(c), where the sizes of the child volumes are large compared to the parent focal volume. In the 60-70\% pressure bin, the cumulative size of the child voxels was 13\% in patient 1 and 9\% in patient 3. However, patient 2 had no voxels in the child volumes above 60\% of the global peak pressure, which can also be seen as a lower degree of splitting in Figure \ref{fig:focal_point_3D}(b). At even higher pressures, at 70-80\% of the global peak pressure, only patient 1 had voxels in the child volumes, with a cumulative size of approximately 5\% of the parent focal point. Above 80\% of the global peak pressure none of the patients had voxels in child regions. The total volumes of the child regions with respect to the parent regions were 46\%, 23\% and 90\% for patients 1-3, respectively. In other words, patient 3 had the highest degree of focal splitting. These data suggest that undesired heating effects might occur at child focal points due to focal point splitting.

\subsection*{Temperature evolution and thermal dose}

\begin{figure}[b!]
    \centering
    \subfigure[]
    {
        \includegraphics[width=0.46\columnwidth]{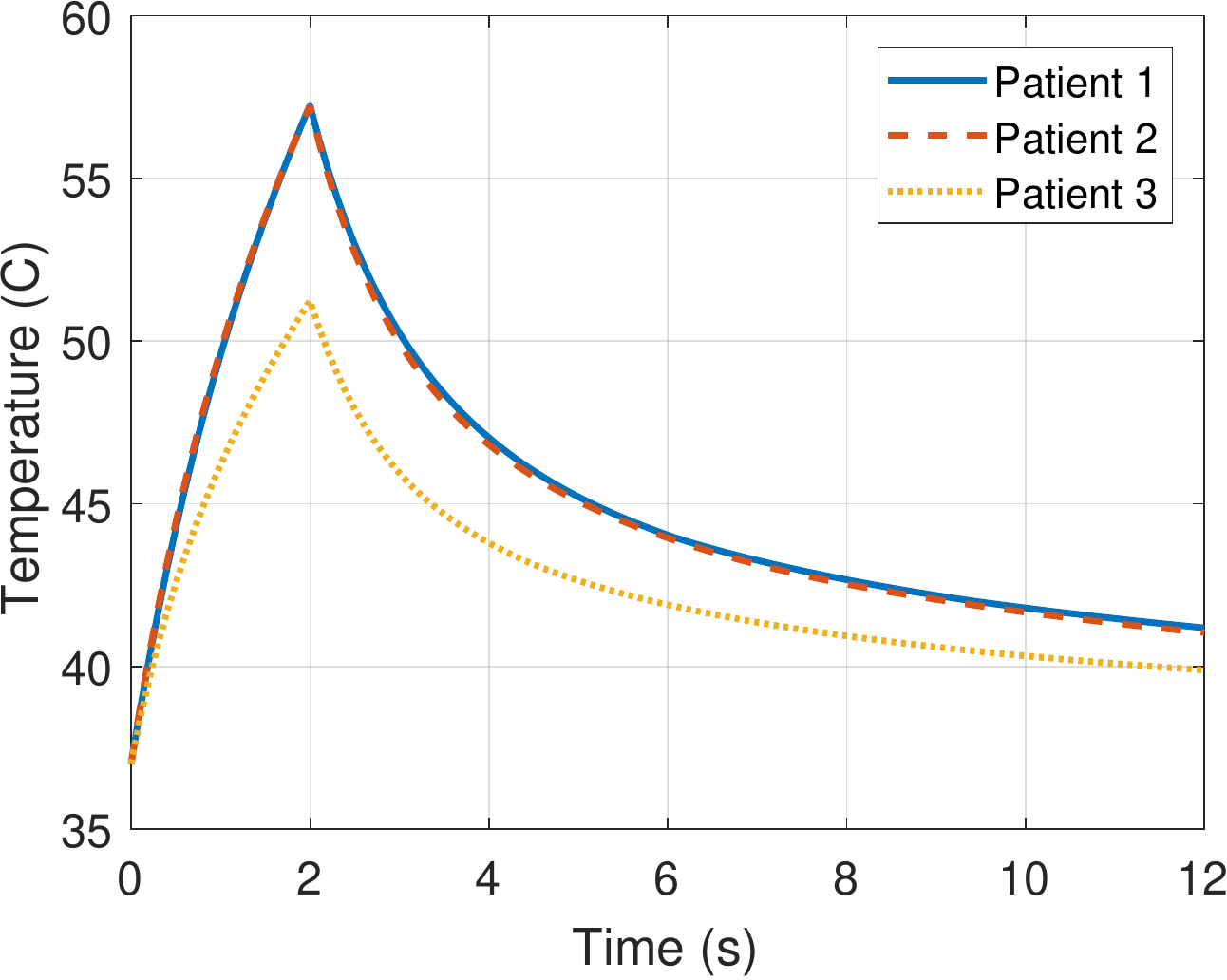}
    }
    \subfigure[]
    {
        \includegraphics[width=0.46\columnwidth]{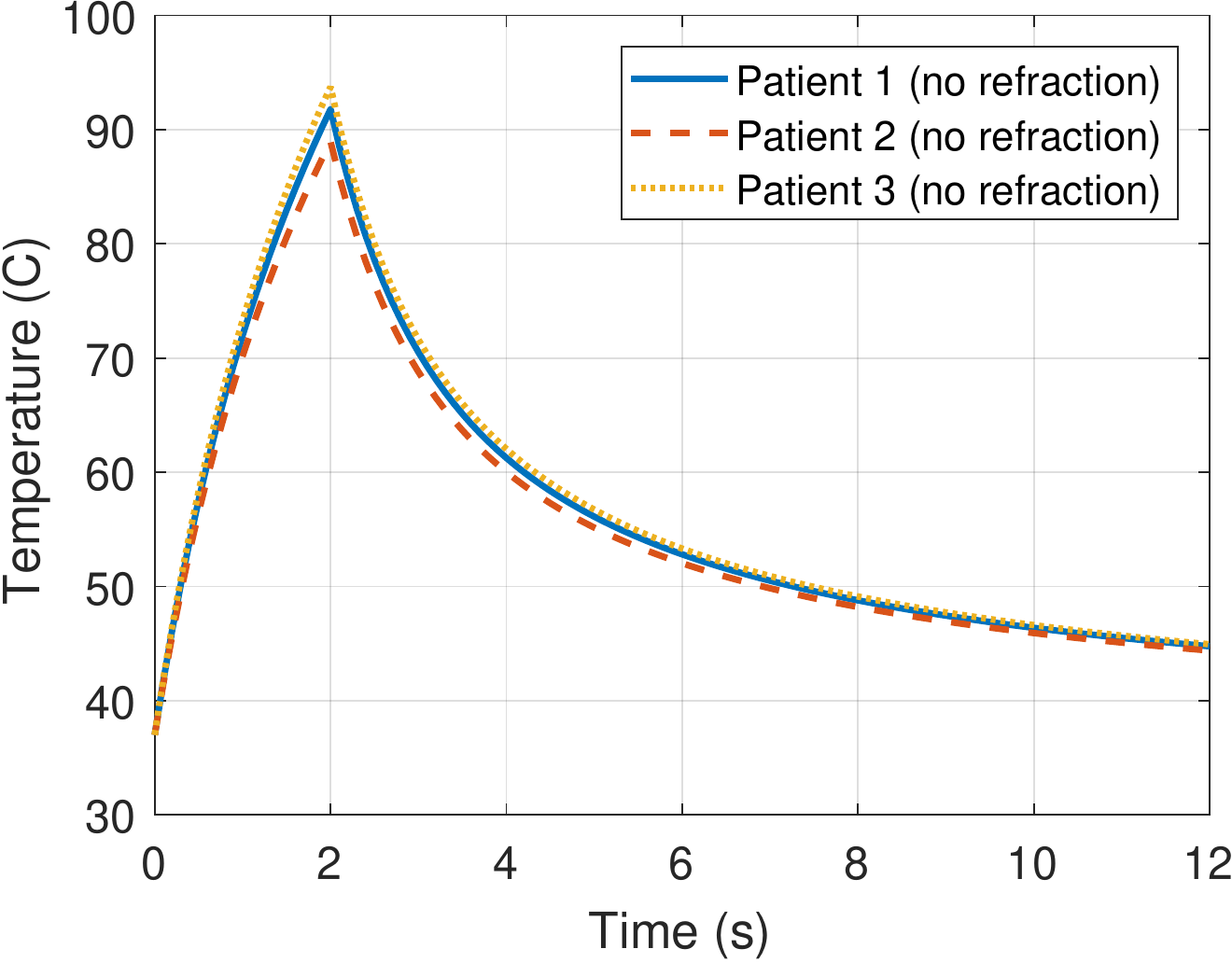}
    }
    \caption{Evolution of maximum temperature with time during a 2-second sonication in the kidney of all three patients (a) with refraction; and (b) without refraction (i.e., constant sound speed in all tissues).}
    \label{fig:temperature_time_patient_comparison}
\end{figure}

The evolution of the maximum temperature during a 2-second sonication in the three patients with and without refraction are shown in Figures \ref{fig:temperature_time_patient_comparison}(a) and (b), respectively. For the simulations with refraction (see Figure \ref{fig:temperature_time_patient_comparison}(a)), the temperature evolution in patients 1 and 2 follow similar trends with respective peak temperatures of 57.2 and 57.2~$^{\circ}$C at the end of the sonication. In patient 3, however, the peak temperature is 51.2~$^{\circ}$C, a 30\% decrease in temperature elevation (from 37~$^{\circ}$C) compared to the other two patients. This is most likely due to the higher degree of focal splitting. On average the peak temperature at the end of the sonication was 55.2~$^{\circ}$C when refraction was included (see Table \ref{tab:thermal_results} for summary).

\begin{table}[t!]
  	\centering
  	\caption{Thermal simulation results}
  	\begin{adjustbox}{max width=\columnwidth}
  	\begin{threeparttable}
    \begin{tabular}{lcccccc}
    \hline
     		& Peak & \multicolumn{2}{c}{Peak temp. shift} & \multicolumn{3}{c}{240 CEM$_{43 ^{\circ}\mathrm{C}}$ size} \\
     		& temperature & Axial	& Radial	& Length & Width & Volume \\
     		& ($^{\circ}$C)	& \multicolumn{2}{c}{(mm)} & \multicolumn{2}{c}{(mm)}	 & (mm$^{3}$) \\	
     \hline
    Patient 1 & 57.2  & 2.0   & 2.4   & 4.6   & 1.5   & 3.6 \\
    Patient 2 & 57.2  & 1.3   & 1.6   & 5.0   & 1.2   & 3.8 \\
    Patient 3 & 51.2  & 2.8   & 1.6   & 1.0   & 0.2   & 0.2 \\
    \hline
    \textbf{Average} & \textbf{55.2} & \textbf{2.0} & \textbf{1.9} & \textbf{3.5} & \textbf{1.0} & \textbf{2.5} \\
    \textbf{SD} & \textbf{2.8} & \textbf{0.6} & \textbf{0.4} & \textbf{1.8} & \textbf{0.6} & \textbf{1.7} \\
    \hline
    Patient 1$^{*}$ & 91.8  & 1.7   & 0.3   & 8.5   & 2.7   & 30.0 \\
    Patient 2$^{*}$ & 88.9  & 1.3   & 0.3   & 8.4   & 2.6   & 27.0 \\
    Patient 3$^{*}$ & 93.7  & 0.9   & 0.3   & 8.5   & 2.9   & 30.3 \\
    \hline
    \textbf{Average} & \textbf{91.5} & \textbf{1.3} & \textbf{0.3} & \textbf{8.5} & \textbf{2.7} & \textbf{29.1} \\
    \textbf{SD} & \textbf{2.0} & \textbf{0.3} & \textbf{0.0} & \textbf{0.0} & \textbf{0.1} & \textbf{1.5} \\
    \hline
    \end{tabular}
    \begin{tablenotes}
  	\item[*] \footnotesize simulation without refraction effects
  	\end{tablenotes}
  	\end{threeparttable}
    \end{adjustbox}
  \label{tab:thermal_results}
  \vspace{-0.3cm}
\end{table}

The simulations without refraction (see Figure \ref{fig:temperature_time_patient_comparison}(b)) show similar peak temperatures (91.8, 88.9 and 93.7~$^{\circ}$C) in all three patients. Here the peak temperatures at the end of the sonication are significantly higher when compared to the sonications with refraction and the variation is small. This is consistent with the lack of focal splitting and small fluctuations in $I_{\mathrm{SPTA}}$ when refraction was neglected. On average the peak temperature at the end of the sonication was 91.5~$^{\circ}$C, which is approximately 36~$^{\circ}$C higher than with refraction, i.e., a three-fold temperature rise.

\begin{figure}[htbp!]
    \centering
    \subfigure[]
    {
        \includegraphics[height=6cm]{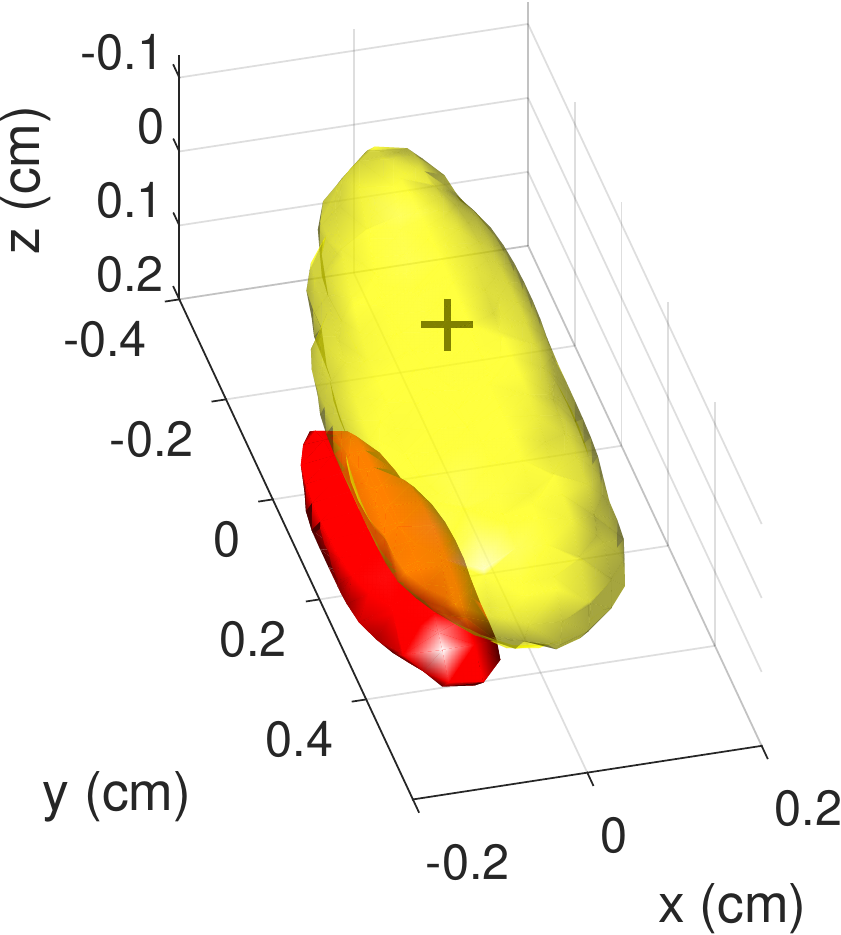}
    }
    \subfigure[]
    {
        \includegraphics[height=6cm]{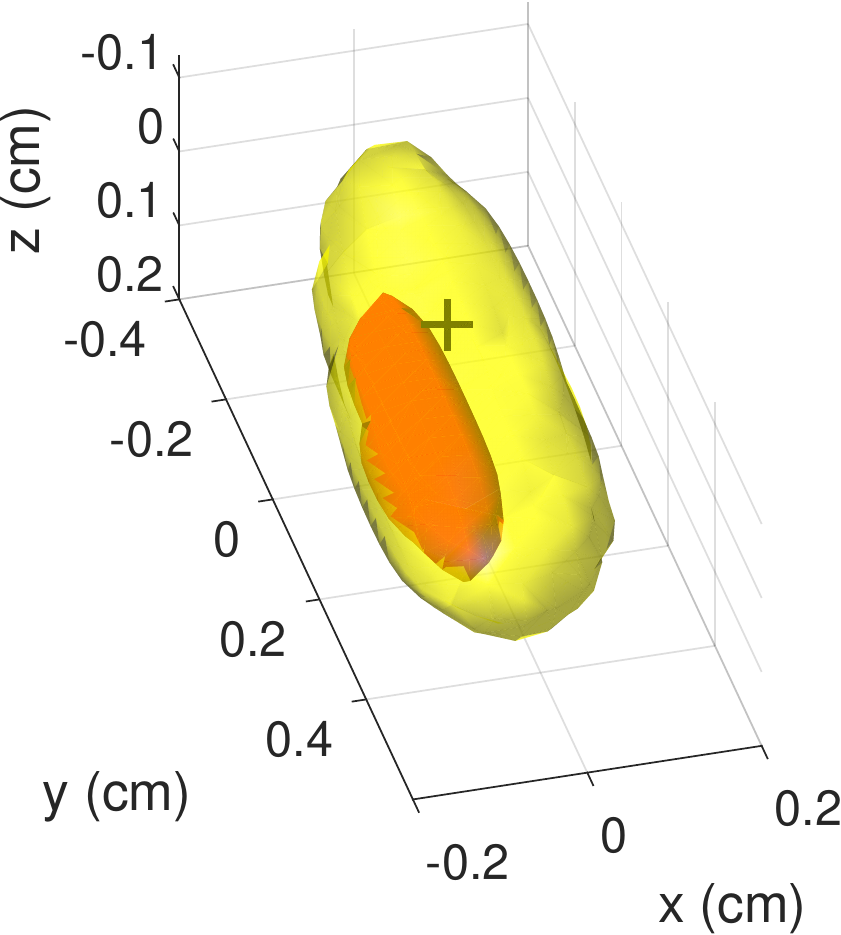}
    }
    \subfigure[]
    {
        \includegraphics[height=6cm]{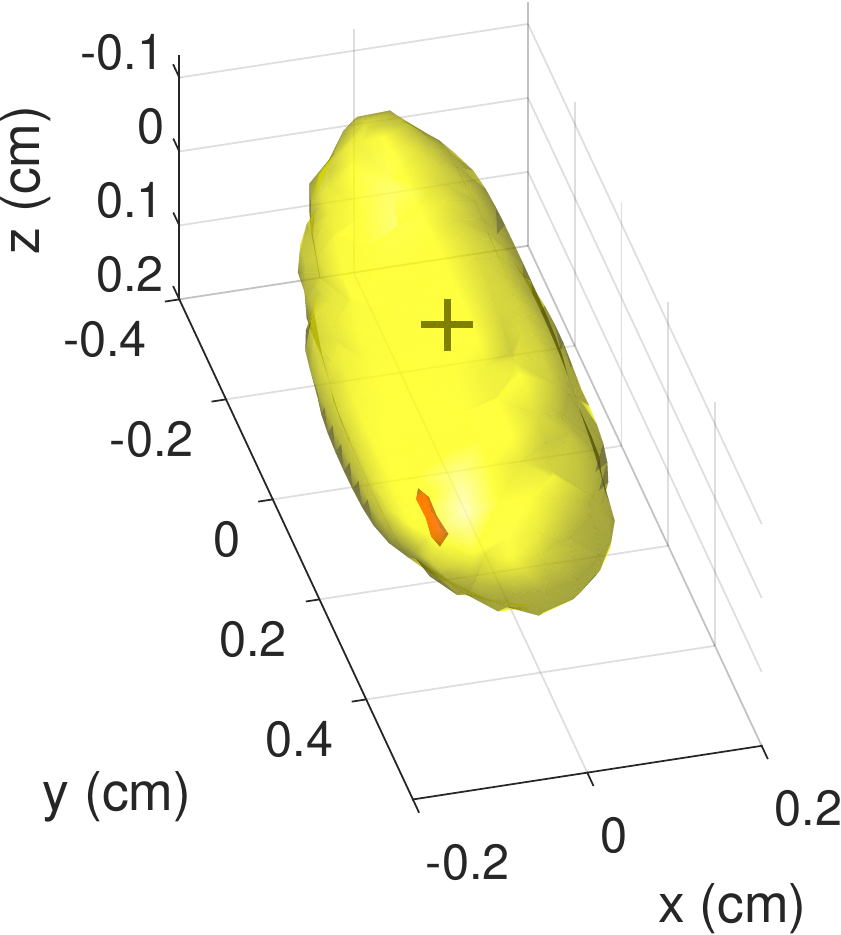}
    }
    \caption{(a)-(c) Thermal dose volumes in the kidney after 2-second sonications for patients 1-3, respectively. The 240 cumulative equivalent minutes at 43~$^{\circ}$C (CEM$_{43 ^{\circ}\mathrm{C}}$) thermal dose volumes with refraction are shown with red isosurfaces and the volumes without refraction with transparent yellow isosurfaces. The target focal point is marked with a black cross.}
    \label{fig:thermal_dose_3D}
\end{figure}

The change in peak temperature, however, is not the complete story when it comes to thermal ablation. Treatment is desired over a volume and the location and the extent of the volume is important. Figures \ref{fig:thermal_dose_3D}(a)-(c) show the 240 CEM$_{43 ^{\circ}\mathrm{C}}$ isosurfaces for each sonication both with (red) and without (yellow) refraction.

It can be seen that adding refraction decreased the treated volume significantly (to almost zero for patient 3) and resulted in shifts of the volume away from the target. Not only are the volumes significantly larger when refraction is absent, they are also evenly located around the target focal point. The size differences are also apparent when comparing the values in Table \ref{tab:thermal_results}. On average the thermal dose volume with refraction was 2.5 mm$^{3}$. Without refraction the average volume was 29.1 mm$^{3}$ which is 11 times larger. Therefore, it is evident that focal point splitting is significantly affecting the creation of thermal dose in the kidney.

Table \ref{tab:thermal_results} gives numerical data for the changes in temperature and location of the thermal simulations. The average shifts in the peak temperature were 2.0 mm in the axial and 1.9 mm in the radial directions with refraction. These are comparable to the corresponding shifts in the peak pressure of 2.1 mm and 1.4 mm. Without refraction the shifts were reduced to 1.3 mm and 0.3 mm again comparable to the 1.2 mm and 0.3 mm shifts in the peak pressure locations.

\subsection*{Tissue property variability}

\begin{figure}[b!]
    \centering
    \subfigure[]
    {
        \includegraphics[width=0.46\columnwidth]{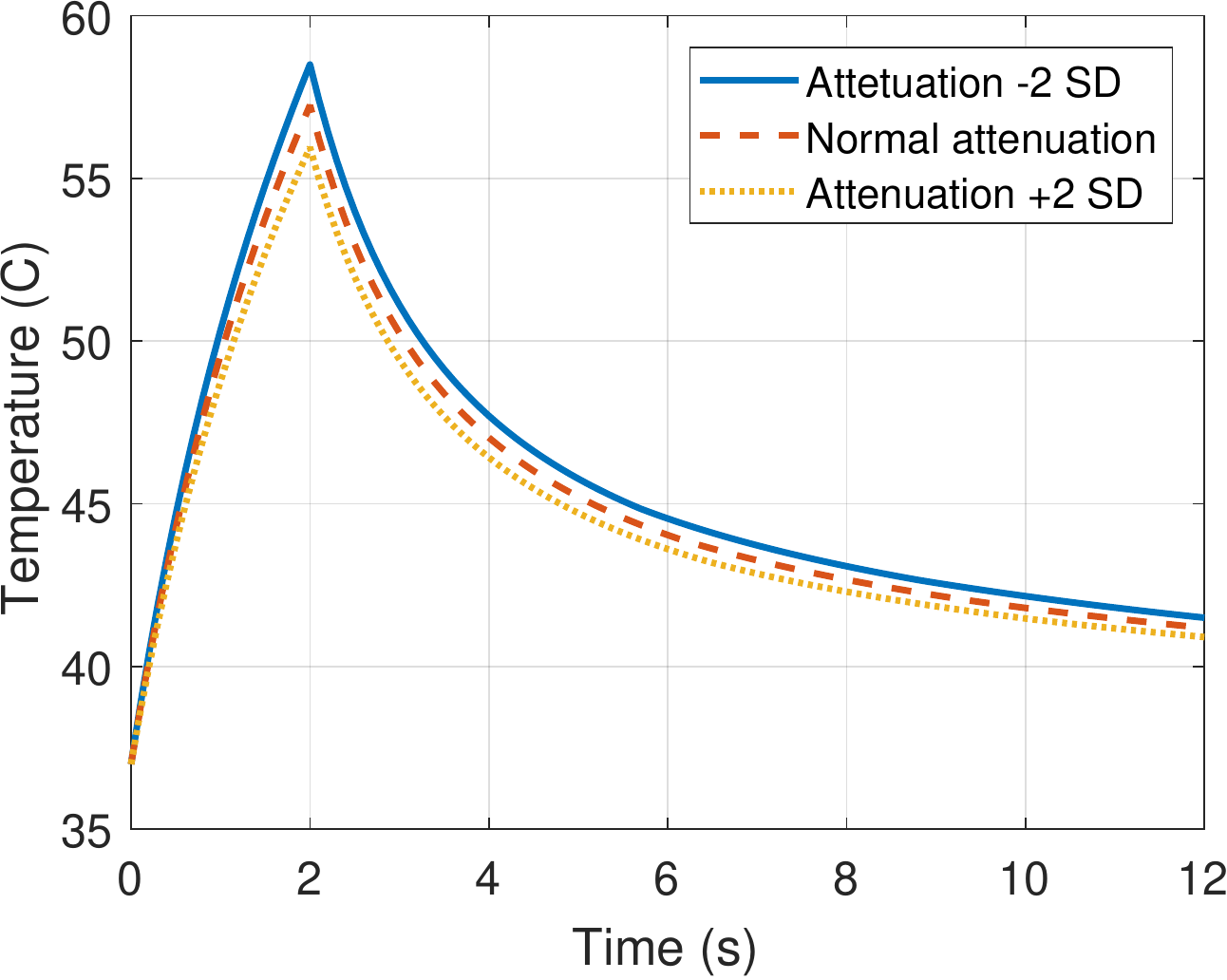}
    }
    \subfigure[]
    {
        \includegraphics[width=0.46\columnwidth]{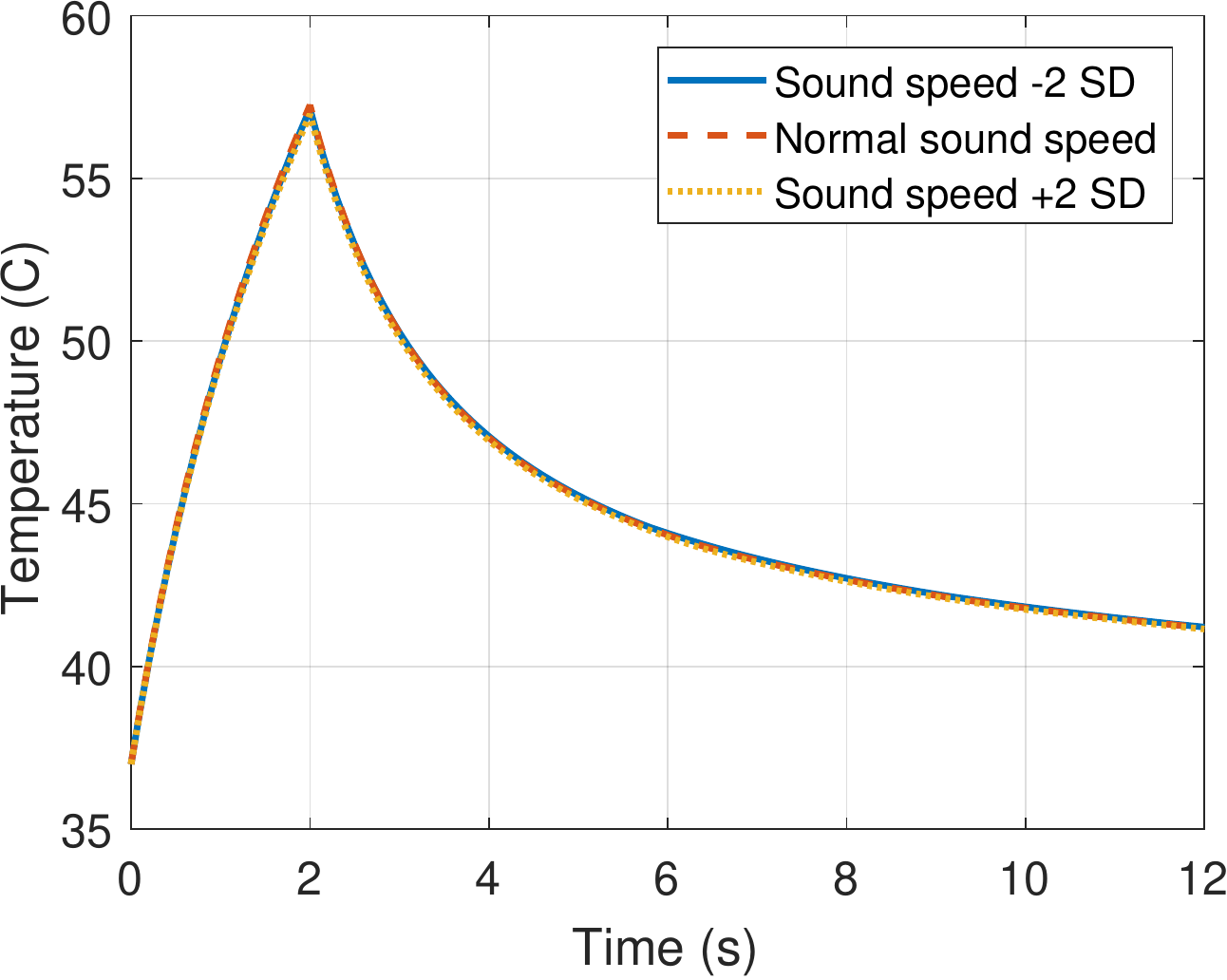}
    }
    \\
    \subfigure[]
    {
        \includegraphics[width=0.46\columnwidth]{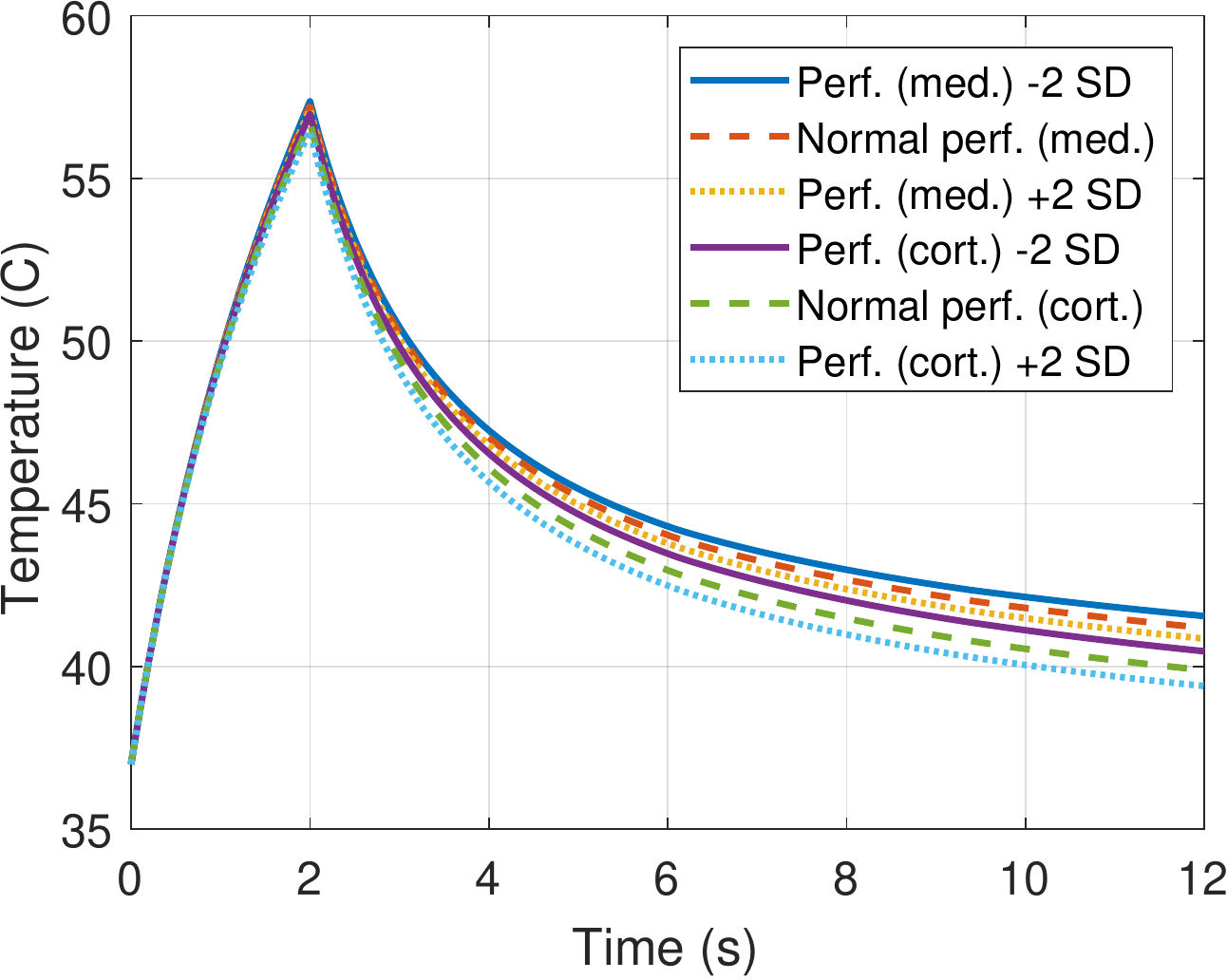}
    }
    \subfigure[]
    {
        \includegraphics[width=0.46\columnwidth]{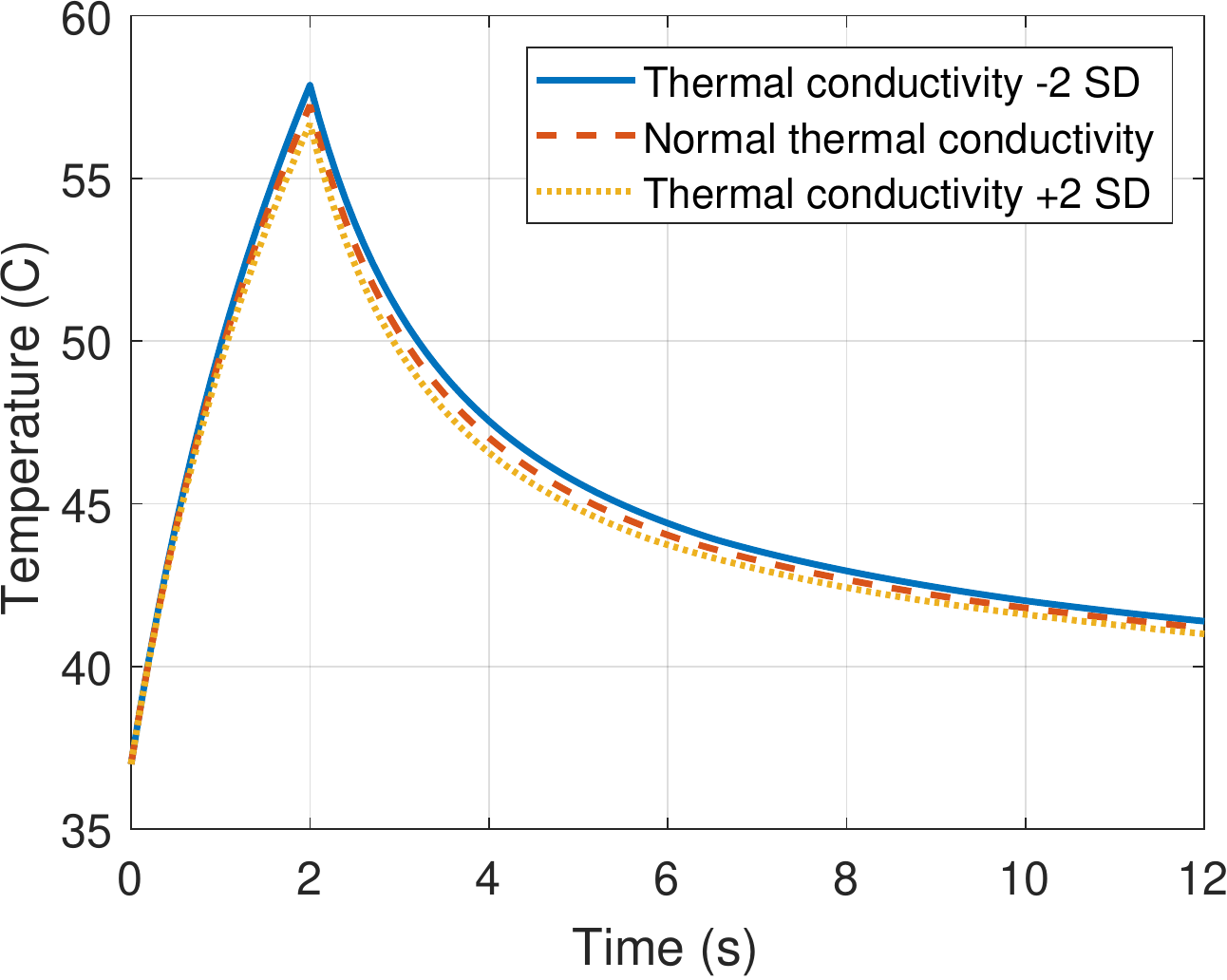}
    }
    \caption{Evolution of maximum temperature with time during a 2-second sonication in the kidney of patient 1 with (a) attenuation, (b) sound speed, (c) perfusion and (d) thermal conductivity of the kidney changing by $\pm$2 standard deviations (SD).}
    \label{fig:temperature_time_comparison}
\end{figure}

The evolution of the maximum temperature, when changing the attenuation, sound speed, perfusion and thermal conductivity of the kidney by $\pm$2 SD in patient 1, are shown in Figures \ref{fig:temperature_time_comparison}(a)-(d), respectively.
Increasing attenuation from 1.00 to 1.24 dB/MHz$^{1.1}$/cm (see Figure \ref{fig:temperature_time_comparison}(a)) results in slightly lower heating with a peak temperature of 56.0~$^{\circ}$C at the end of the sonication when compared to the value of 57.3~$^{\circ}$C with `normal' attenuation. The decrease in attenuation results in a slightly higher peak temperature of 58.5~$^{\circ}$C. These changes are relatively small because the total loss due to attenuation of tissue layers in the ultrasound pathway is the main factor reducing the intensity while the penetration depth in the kidney is short. When the sound speed is changed by $\pm$10 m/s (see Figure \ref{fig:temperature_time_comparison}(b)), no significant differences in the peak temperature are seen. This suggests that the changes in sound speed of the kidney do not result in significant differences in focal point splitting. The primary source of splitting must therefore be caused by the compounded effects of refraction from all the tissue layers in front of the transducer. 

Decreasing perfusion by 8.3 kg/m$^{3}$/s in medulla (see Figure \ref{fig:temperature_time_comparison}(c)) resulted in a peak temperature of 57.4~$^{\circ}$C, which is 0.1~$^{\circ}$C higher than with the normal medulla perfusion value. Increasing medulla perfusion by the same amount resulted in 0.1~$^{\circ}$C lower peak temperature of 57.1~$^{\circ}$C. In the cortex, temperature with the normal perfusion reached 56.7~$^{\circ}$C while temperature values of 56.5 and 57.0~$^{\circ}$C were achieved by increasing and decreasing cortex perfusion by 18.3 kg/m$^{3}$/s, respectively. The perfusion rate of renal cortex is approximately five-fold higher \citep{roberts1995renal}, which resulted in 3\% lower temperature rise in patient 1. This suggests that perfusion is not a significant parameter affecting the heating efficacy in the kidney with short sonication durations. The decrease in thermal conductivity (see Figure \ref{fig:temperature_time_comparison}(d)) resulted in a slightly higher peak temperature of 57.9~$^{\circ}$C when compared to the normal value of 57.3 and 56.7~$^{\circ}$C with increased thermal conductivity.

\subsection*{Phase aberration}

\begin{figure}[b!]
    \centering
    \subfigure[]
    {
        \includegraphics[width=0.3\textwidth]{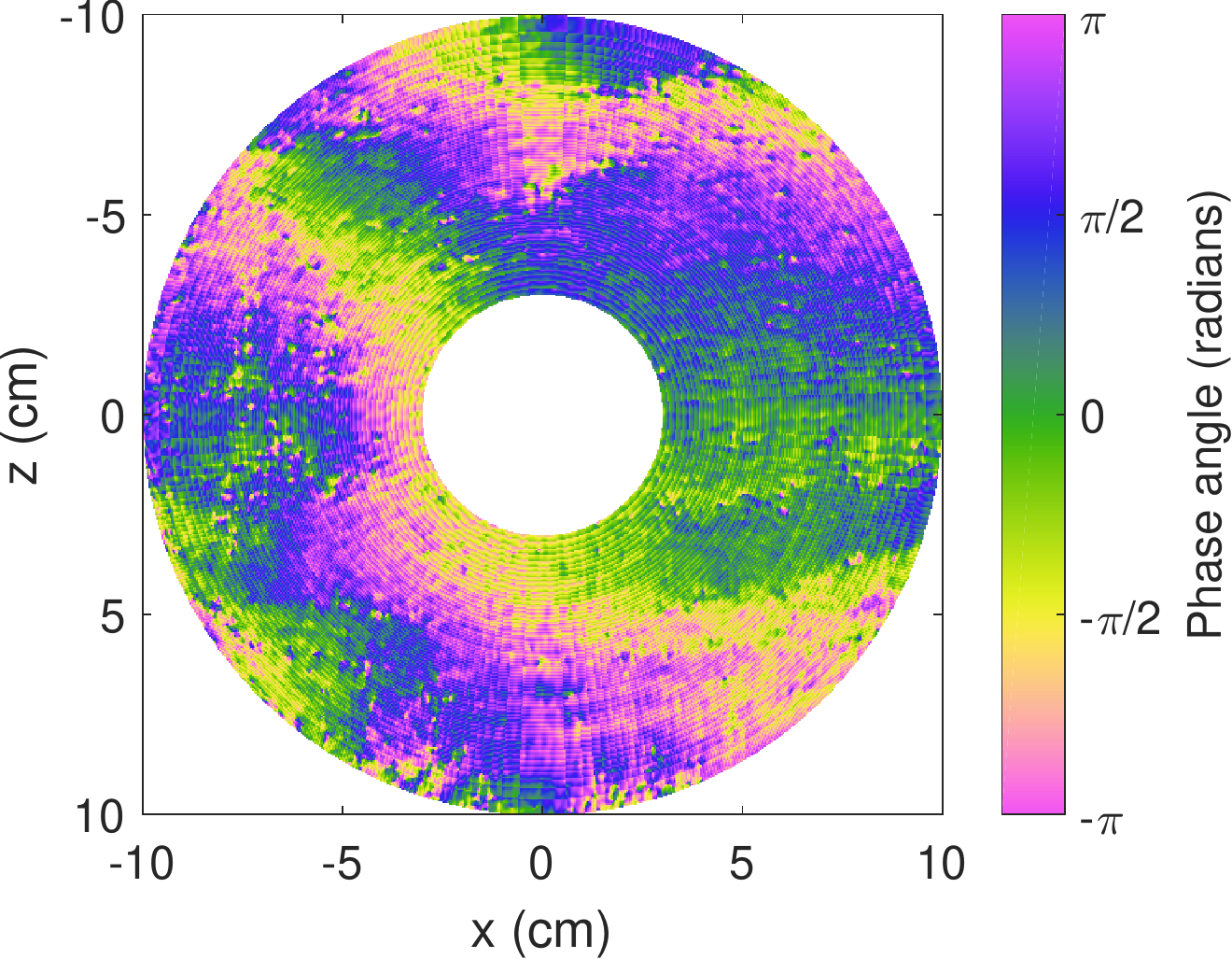}
    }
    \subfigure[]
    {
        \includegraphics[width=0.3\textwidth]{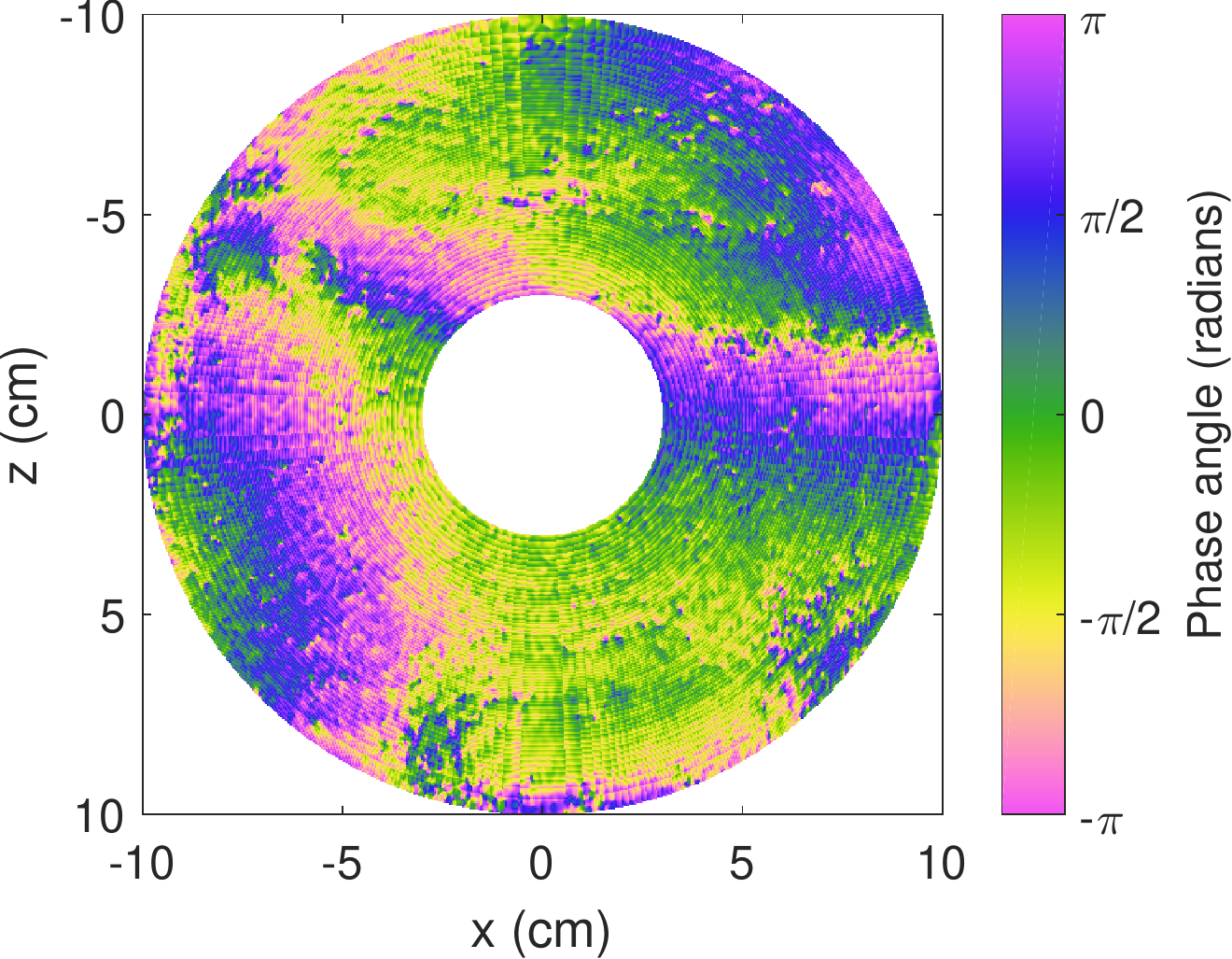}
    }
    \subfigure[]
    {
        \includegraphics[width=0.3\textwidth]{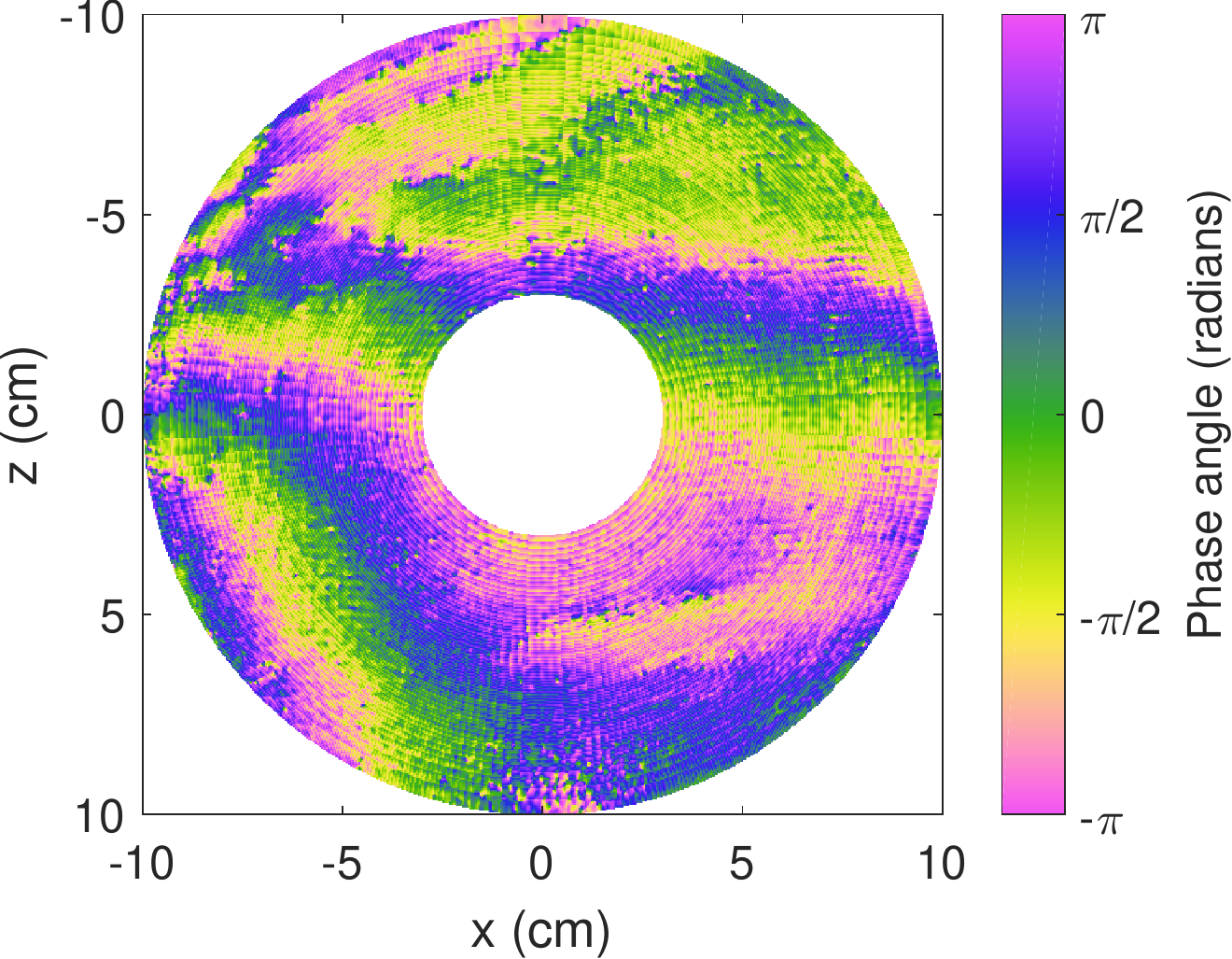}
    }
    \\
    \subfigure[]
    {
        \includegraphics[width=0.3\textwidth]{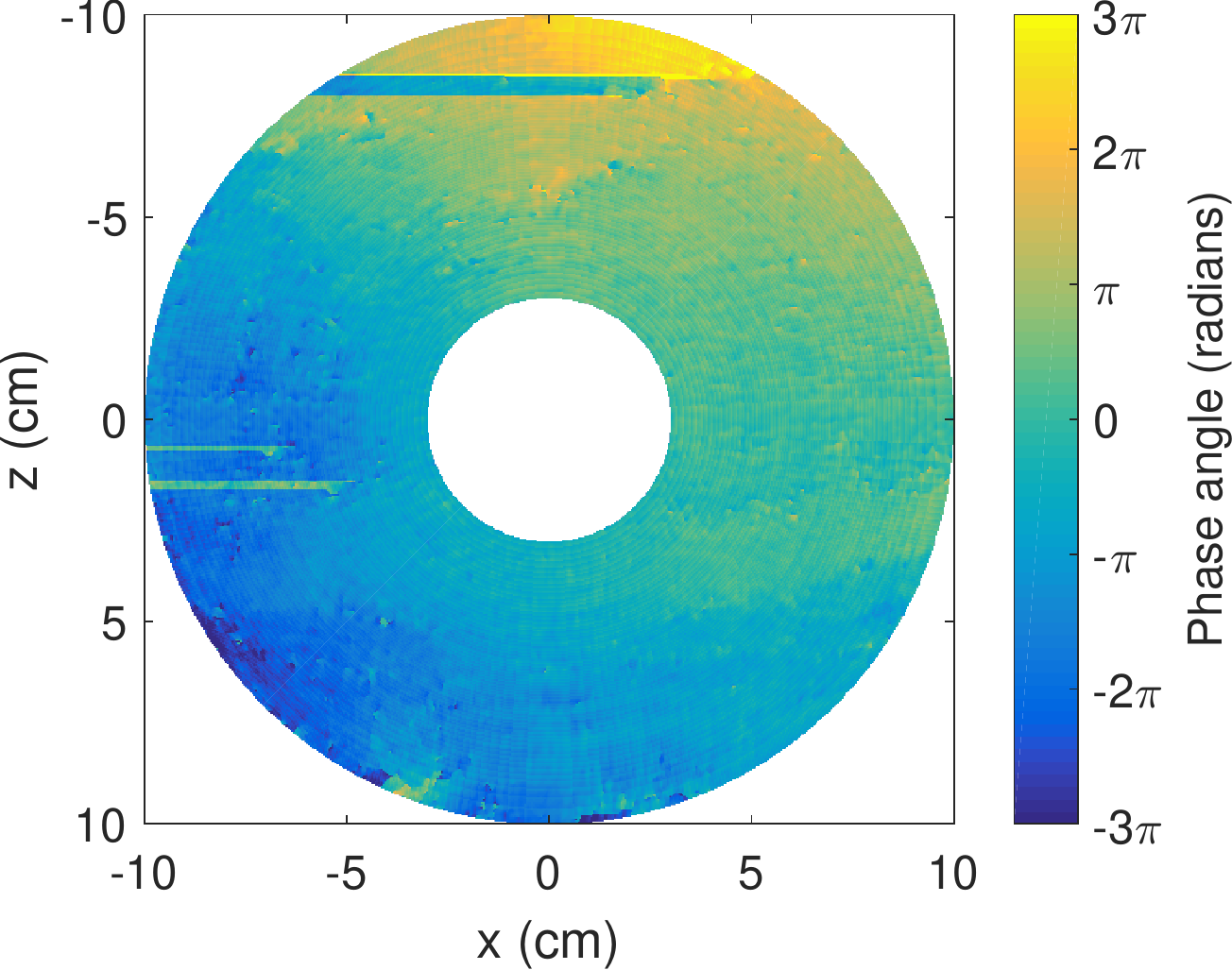}
    }
    \subfigure[]
    {
        \includegraphics[width=0.3\textwidth]{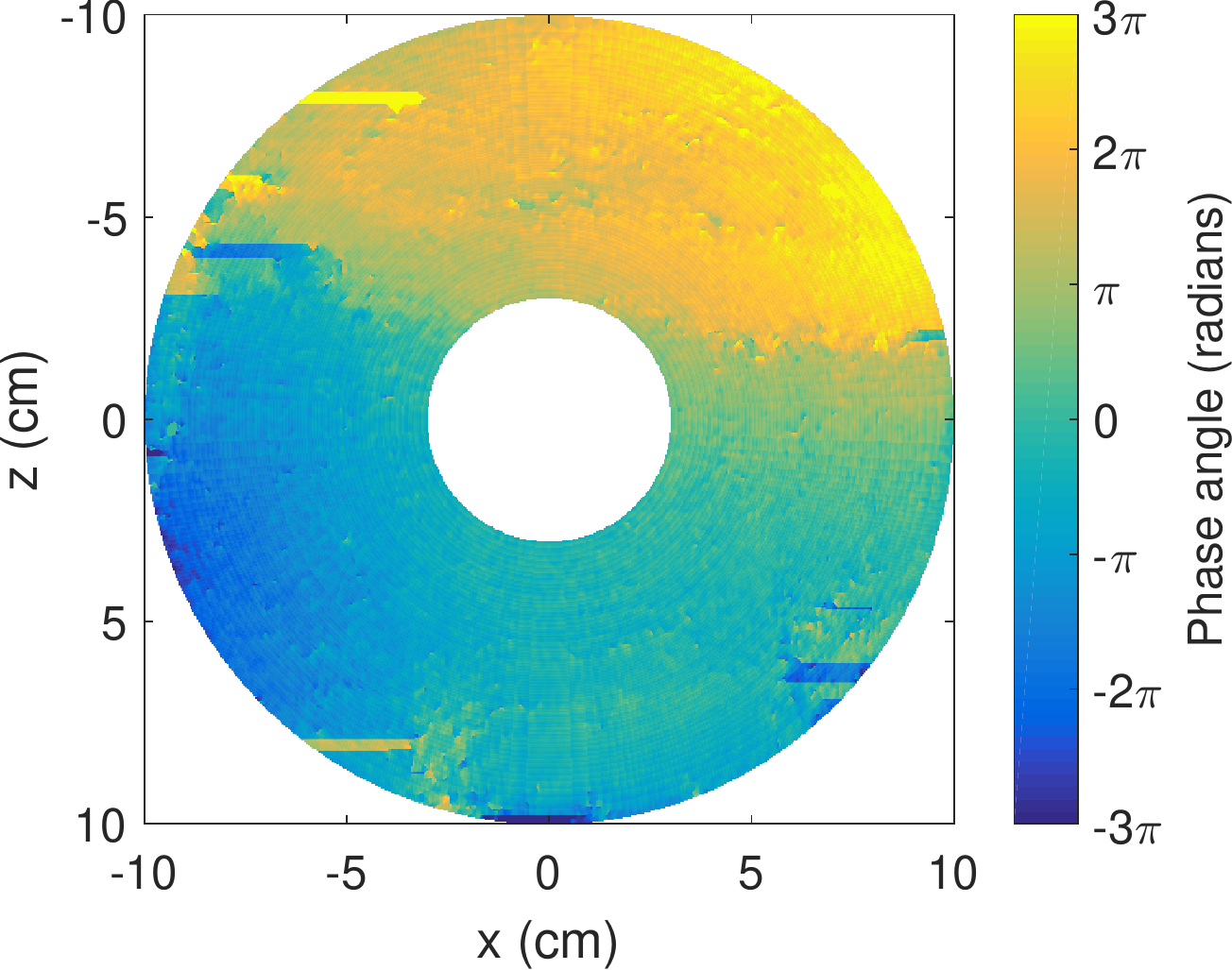}
    }
    \subfigure[]
    {
        \includegraphics[width=0.3\textwidth]{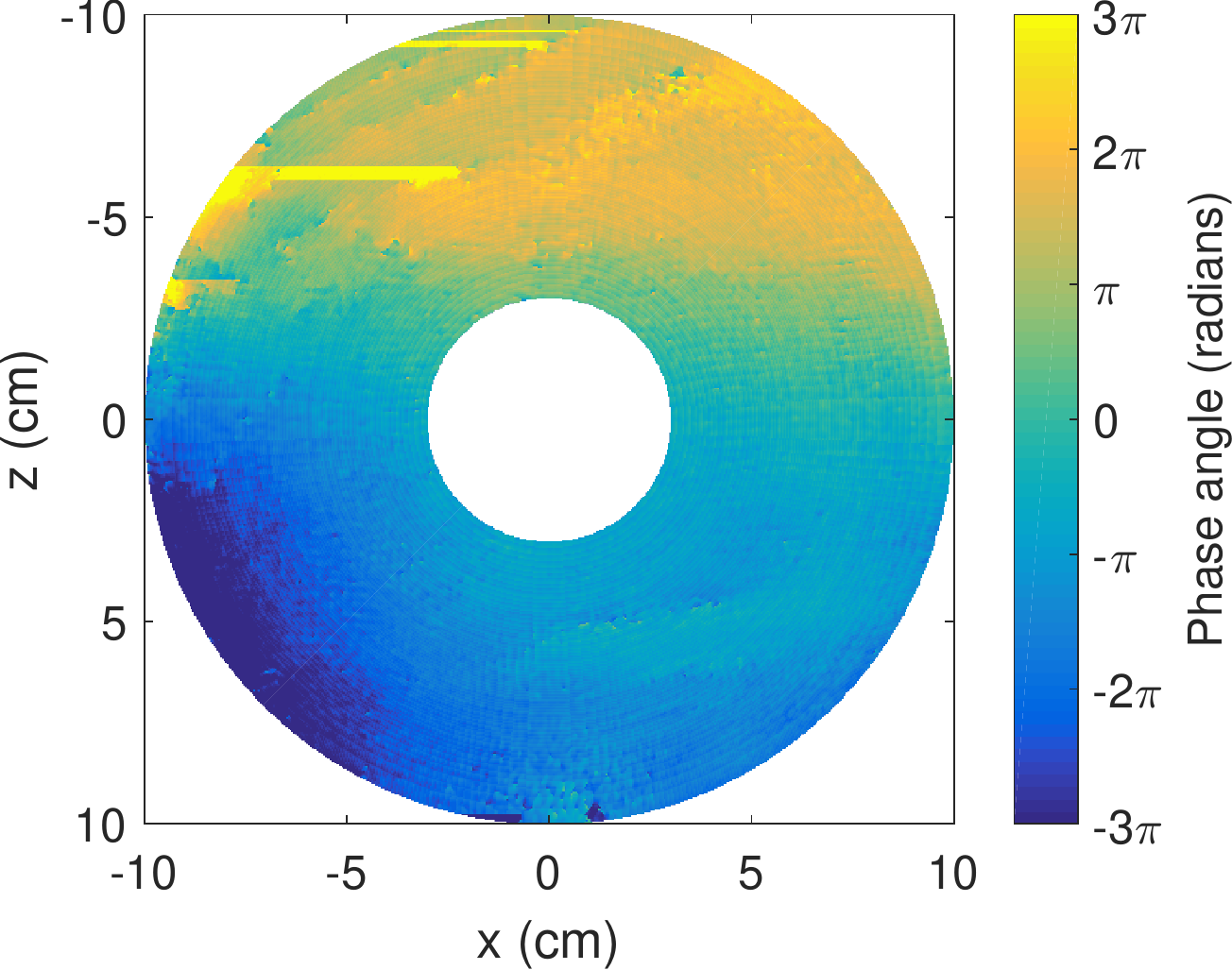}
    }
    \\
    \subfigure[]
    {
        \includegraphics[width=0.3\textwidth]{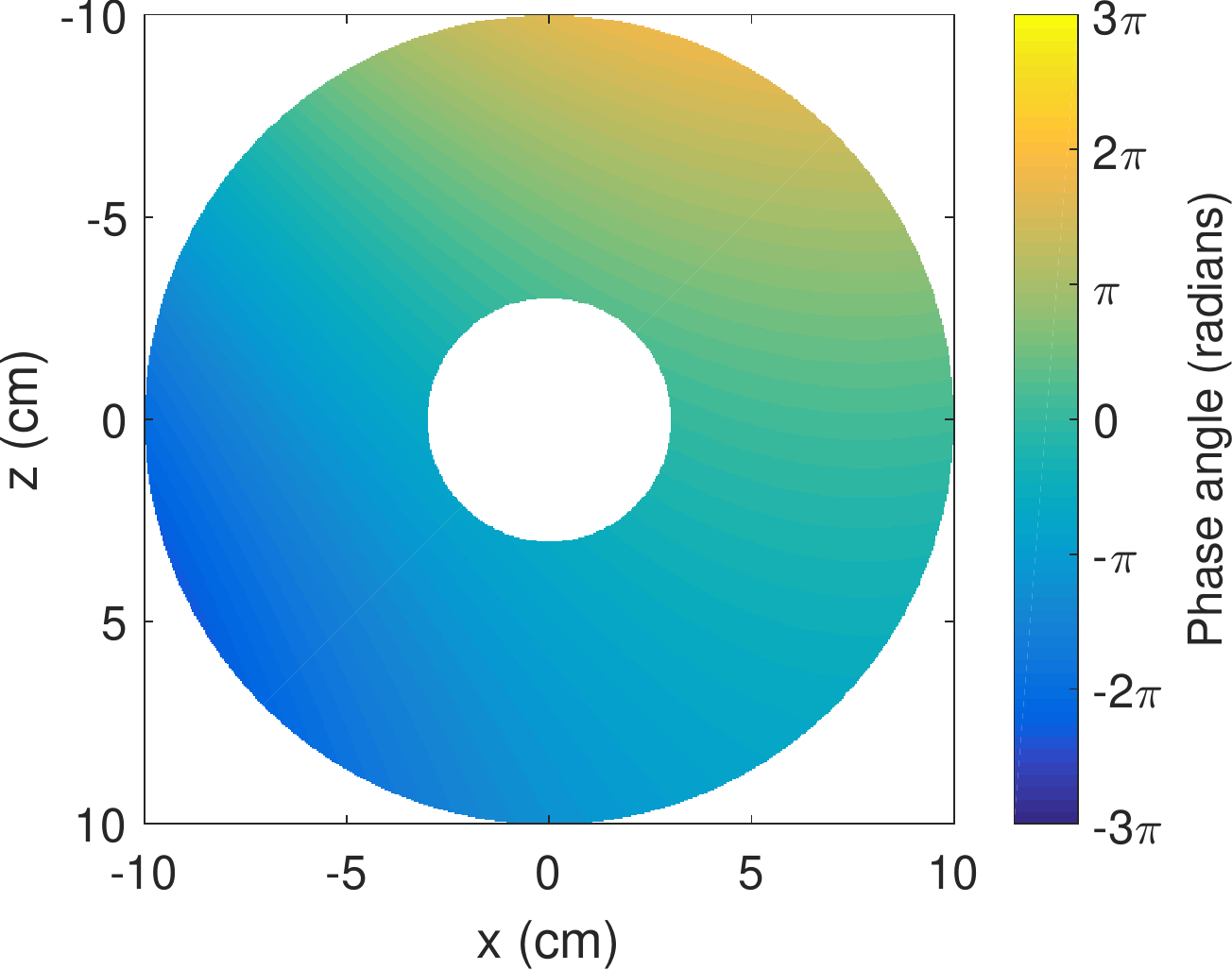}
    }
    \subfigure[]
    {
        \includegraphics[width=0.3\textwidth]{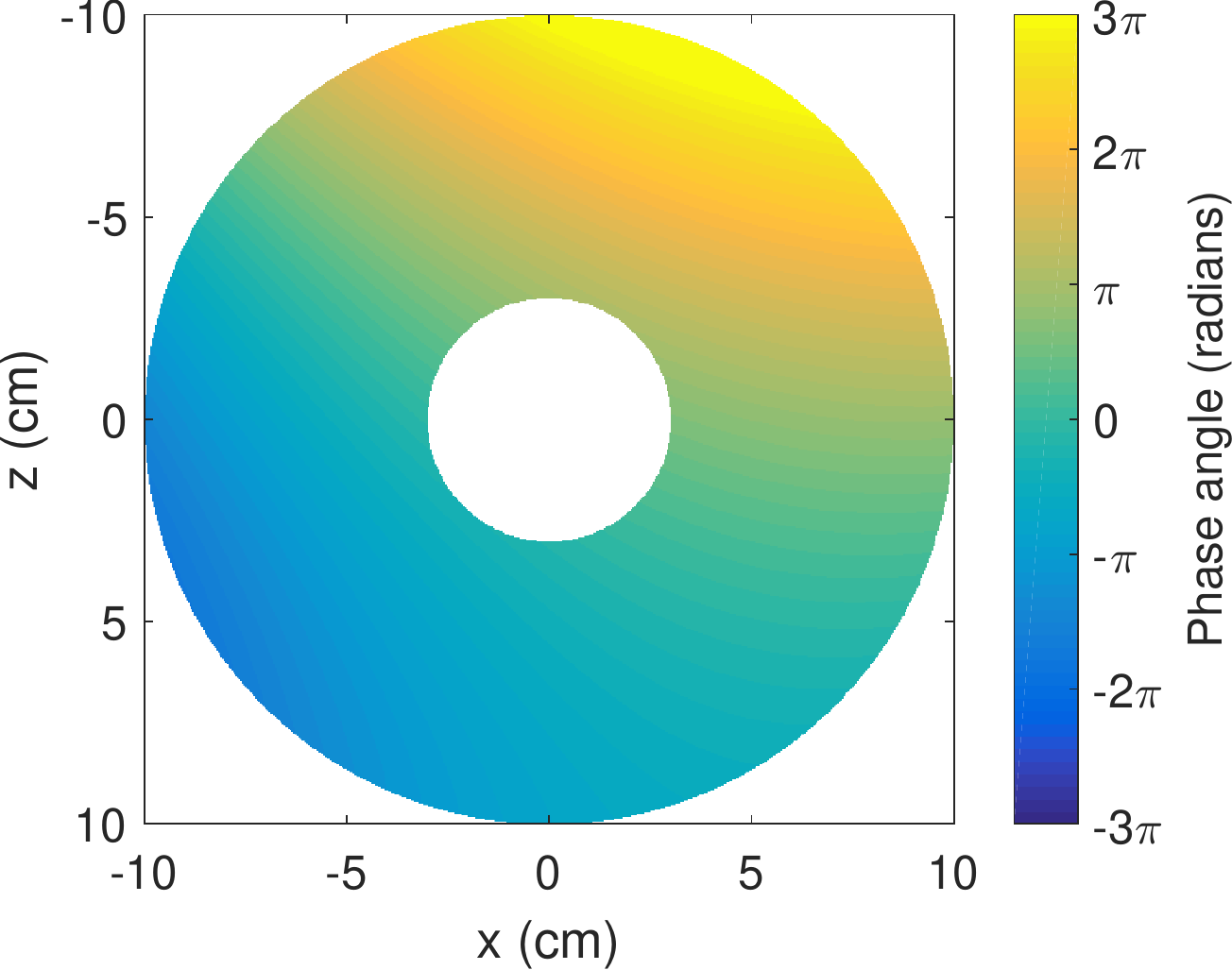}
    }
    \subfigure[]
    {
        \includegraphics[width=0.3\textwidth]{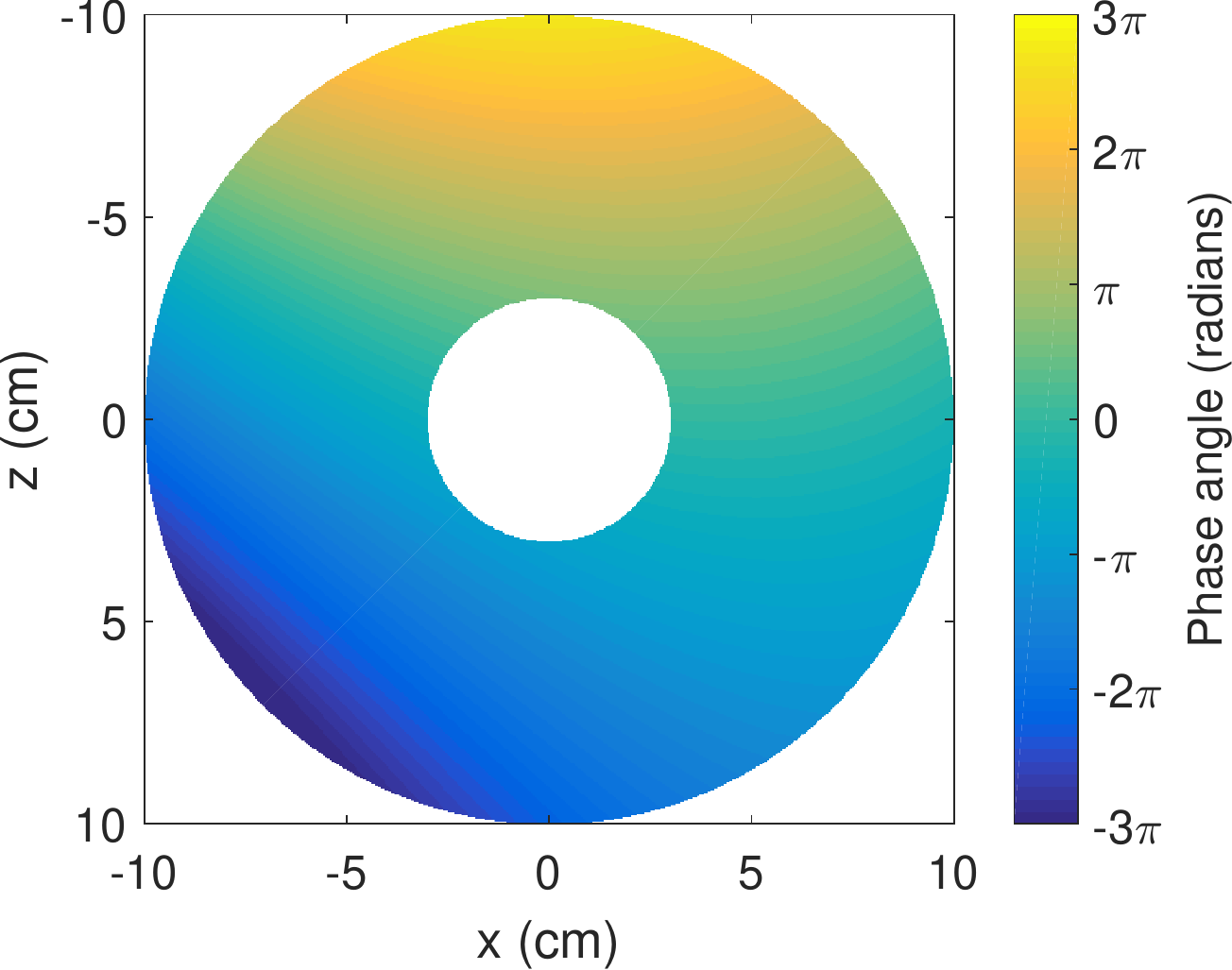}
    }
    \caption{(a)-(c) Wrapped, (d)-(f) unwrapped and (g)-(i) fitted second order polynomial phase shifts for patients 1-3, respectively. The therapeutic transducer was used as a receiver for an acoustic point source located at the geometric focus.}
    \label{fig:aberration}
\end{figure}

The simulations suggest that refraction has a dramatic effect on the desired heating in the kidney. The effects of refraction can be mitigated by adjusting the phase of the source using the principle of time-reversal \citep{fink1992time}. Using this technique, the aberration simulations were conducted with the same transducer positions, acoustic parameters and tissue parameters as in the ultrasound pressure simulations, but in a reverse manner. In this case, the therapeutic transducer was used as a receiver and an acoustic point pressure source was placed at the geometric focus of the transducer (i.e., inside the kidney of the patient). The acoustic point source was set to transmit a continuous wave at 0.95 MHz. The simulations were run using a computational grid of 640 $\times$ 640 $\times$ 640 grid points which supported harmonic frequencies up to 2 MHz. PMLs were used on the edges of the grid. The temporal resolution of the simulations was set to 15.99 ns which gave 17329 time steps per simulation with 277 $\mu$s simulation duration. The simulations were run using 320 computing cores with an average wall-clock time of 5.5 hours per simulation.

For data analysis, three cycles of the ultrasound pressure waveforms at the surface of the therapeutic transducer were saved for each grid point. The phase shifts of these ultrasound pressure waveforms were then calculated using the DFT at the fundamental frequency (0.95 MHz) and projected on the transducer plane for visualisation. The phase shifts were then unwrapped using a two-dimensional phase unwrapping algorithm \citep{goldstein1988satellite, ghiglia1998two} and the unwrapped values were similarly projected on the transducer face.

The phase shifts obtained directly from the DFT for each patient are presented in Figures \ref{fig:aberration}(a)-(c) for patients 1-3, respectively. The phase shifts are presented using a cyclic colour map (i.e., phase shifts of $-\pi$ and $\pi$ are the same colour) which allows visualisation of the areas with similar magnitude phase shifts. Areas of similar magnitude phase shifts are seen crossing the width of the transducer in the radial direction. Furthermore, a `wave-like' behaviour is seen where the phase shifts increase and decrease subsequently when moving towards the upper right corner of the transducer.

The unwrapped phase shifts are presented in Figures \ref{fig:aberration}(d)-(f) for patients 1-3, respectively. The phase shifts follow a smooth trend which starts from the lower left part of the transducer and increases towards the upper right corner in all three patients. The areas with negative phase shifts are on the lower left section while the positive phase shifts are located on the upper right section. Some artifacts appear as horizontal lines in the areas where the algorithm was unable to resolve the underlying phase shift, however, the behaviour suggest that a function representation of the spatial phase shifts is possible due to the relatively smooth transition.

In order to quantify the phase shifts on the transducer face, a second order polynomial function was fitted to the unwrapped phase data in the form:
\begin{align}
\label{eq:phase_shift}
\varphi(x, z)\ \text{(radians)} =&\ p_{00} + p_{10}x + p_{01}z \nonumber \\
& + p_{20}x^{2} + p_{11}xz + p_{02}z^{2} 
\end{align}
where $\varphi$ is the phase shift at the spatial location ($x$, $z$); and $p_{00}$-$p_{02}$ are the fitting parameters with the corresponding subscripts. The areas of horizontal artifacts were included in the fits due to their relatively small size with respect to the transducer face. The second order polynomial fits to the unwrapped phase data according to are shown in Figures \ref{fig:aberration}(g)-(i) for patients 1-3, respectively. The polynomial fitting parameters together with the corresponding coefficients of determination $R^{2}$ are shown in Table \ref{tab:aberration_fit}. In all three patients, the second order polynomial function represents the underlying spatial phase shifts with $R^{2}$ $\ge$ 0.83.

\begin{table}[t!]
  \centering
  \caption{The second order polynomial surface fitting parameters}
  	\begin{adjustbox}{max width=\columnwidth}
    \begin{tabular}{lrrrrrrr}
    \hline
          		& \multicolumn{1}{c}{$p_{00}$}  & \multicolumn{1}{c}{$p_{10}$}  & \multicolumn{1}{c}{$p_{01}$}  & \multicolumn{1}{c}{$p_{20}$}  & \multicolumn{1}{c}{$p_{11}$}  & \multicolumn{1}{c}{$p_{02}$}  & \multicolumn{1}{c}{$R^{2}$} \\
    \hline
    Patient 1  	& $-$0.9749 & 0.3462 	& $-$0.4437	& $-$0.0215 & $-$0.0002 & 0.0147 	& 0.831 \\
    Patient 2	& 1.0920 	& 0.3599 	& $-$0.5993 & $-$0.0198 & $-$0.0140 & 0.0253 	& 0.840 \\
    Patient 3 	& $-$0.5279 & 0.2506 	& $-$0.7787 & $-$0.0291 & 0.0260 	& 0.0131 	& 0.872 \\
    \hline
    \end{tabular}
    \end{adjustbox}
  \label{tab:aberration_fit}
\end{table}

\section*{Discussion}

In this study acoustic and thermal simulations in the kidney have been carried out using realistic patient models. In the acoustic simulations the $I_{\mathrm{SPTA}}$ in the kidney dropped by 11.1 dB (92\%) relative to water, due to a combination of attenuation and refraction. Simulations in the absence of refraction  resulted in a 6.4 dB (77\%) drop in the $I_{\mathrm{SPTA}}$, which can be attributed to attenuation. This implies that refraction accounts for the 4.7 dB difference. In the simulations performed here all the fat in body was segmented into one region and a uniform attenuation of 0.48 dB/cm \citep{mast2000empirical} was applied. \citet{ritchie2013attenuation} measured the attenuation of peri-nephric fat (which surrounds the kidney) and found it to be significantly higher: 1.36 dB/cm. We estimate that incorporating higher values in the peri-nephric regions would result in an extra 0.35 to 1.41 dB of loss as the thickness of peri-nephric fat was approximately 0.4-1.6 cm. This would mean a modest increase in the importance of attenuation on intensity drop but the most significant attenuation losses would still be due to subcutaneous fat and soft tissue in front of the kidney whose thickness were approximately 1.8-2.6 and 3.0-5.0 cm, respectively.

Other potential mechanisms of energy loss should be minimal for the scenarios considered here. The rib cage was avoided by the careful placement of the transducer. Transmission coefficients at tissue interfaces were estimated using plane wave coefficients and found to be: water-fat 99.84\%, fat-soft tissue 99.29\%, soft tissue-fat 99.29\% and fat-kidney 99.41\%. For all the interfaces the estimated intensity transmission is 97.85\% which corresponds to a loss of less than 0.1 dB. This is consistent with the findings by Damianou \citep{damianou2004mri}, who studied the penetration of HIFU in rabbit kidney \textit{in vivo}. They found the ultrasound penetration through muscle-kidney and fat-kidney interfaces to be excellent in a situation where no air bubbles were present. They did report strong reflections only in the case where air spaces existed in between these interfaces, something not included in the model and not anticipated clinically.

The effect of refraction was shown to be important in: (i) reducing the focal intensity, (ii) shifting the location of the focus and (iii) altering the spatial distribution of the intensity. In order to capture the refraction effects it was necessary to have a fully three-dimensional heterogeneous simulation \citep{tolstoy19963}. Focal shifting due to subcutaneous and peri-nephric fat was studied by \citet{ritchie2013attenuation} who found the shift to be approximately 1 mm in both transverse directions. However, in reality the shifts are not only affected by the thickness of the tissue layers but also their geometries. In the simulations reported here the average axial shift was 2.1 mm and the transverse shift 1.4 mm comparable to the results of \citet{ritchie2013attenuation}. The axial shifts are relatively small compared to the average axial focal length of the parent (9.4 mm), however in the transverse direction the shifts are of the same order as the $-$6 dB focal width (1.5 mm), and thus, would result in an offset in the lesion creation. Although these shifts are small compared to a typical renal tumour sizes of several centimetres \citep{remzi2007renal}, it could mean tumour boundaries are not treated properly and if the offset varies across the kidney regions of untreated tissue could result. This motivates monitoring temperature during HIFU treatment to ascertain where ablation occurs, using for example MRI.

In addition to shifting the focus refraction also resulted in the splitting of the focal volume into smaller, diffuse, volumes resulting in lower intensity and hence reduced temperature rises. When focal point splitting was present, the cumulative size of the child focal volumes were found to be between 23-90\% of the parent volume. The highest degree of focal splitting was observed in patient 3, which also had the thickest layer of peri-nephric fat in front of the kidney (1.6 cm). This suggests that the thickness of peri-nephric fat can have a more significant effect on refraction rather than attenuation as suggested previously \citep{ritchie2013attenuation}. The pressure distribution analysis in the child volumes also showed peak pressures reaching up to 80\% of the global peak pressure, which suggest that heating in the child regions will be comparable to the parent volumes. Therefore, focal splitting could potentially cause heating of peripheral areas similarly to focal shifting. Skin heating might also occur if higher pressures are used to achieve greater heating efficacy. However, the use of higher pressures can possibly be avoided by using aberration correction to increase the efficacy of the therapy.

Thermal simulations confirmed the effect of focal splitting on heating patterns. For simulations without refraction the expected ellipsoidal elevated temperature region was observed in all three patients with comparable temperatures and thermal doses present in the target region. When refraction was included the increase in temperature dropped from 55~$^{\circ}$C to 18~$^{\circ}$C. In patients 1 and 2 the volume of tissue that exceed a thermal dose of 240 CEM$_{43 ^{\circ}\mathrm{C}}$ (a typical threshold for ablation) was reduced by a factor of 11 and the location was offset from the target by approximately 2 mm. Notably patient 3, which showed the greatest focal splitting in the acoustic simulations, had a peak temperature rise of approximately 14~$^{\circ}$C and only a small tissue region had a thermal dose that exceeded 240 CEM$_{43 ^{\circ}\mathrm{C}}$. These data suggest that refraction can dramatically reduce heating and that it can result in far more patient variability than attenuation. The later statement is not surprising if one recalls Snell's law to recognise that refraction is sensitive to the angle of incidence and therefore differences in patient geometry will affect refraction.

Inspired by time-reversal \citep{fink1992time}, a strategy to mitigate the effects of refraction was investigated in which a virtual source was placed at the focus and the sound back propagated on to the source plane from which the phase was calculated. A relatively smooth variation in the phase across the transducer was observed which could be modelled by a second order polynomial. These data indicate that by controlling source phase the aberration can be corrected and so the intensity loss and focal splitting due to refraction could be minimised, for example, by using a phased array transducer. The parameters for the phase were patient specific and therefore it would be necessary to do treatment planning calculations on a patient-by-patient basis. In this case a multi-element phased array HIFU transducer should be used \citep{mougenot2012high}. Furthermore, the optimal sizes, locations and number of source elements in a phased array transducer should be specified in order to account for aberration effects and grating lobes \citep{dupenloup1996reduction}. However, the determination of these parameters is beyond the scope of this study, but should be considered in the future research.

Variability in the acoustic and thermal fields would also be expected due to variations in the tissue parameters and therefore simulations were carried out in which the attenuation, sound speed, perfusion and thermal conductivity of the kidney were changed over their physiological range \citep{roberts1995renal, turnbull1989ultrasonic, hasgall2015database}. Attenuation, thermal conductivity and sound speed were all found to have little effect on the results; although only values in the kidney were changed and variations in other layers could have a more significant effect particularly sound speed given the patient sensitivity to refraction. Perfusion did not have a large effect on temperature elevation during short sonication durations, but the cooling rate was noticeably faster in the cortex. These results are consistent with observations made by \citet{chang2004effects}, who found the obstruction of the blood flow to increase the size of the created thermal lesions in kidneys during long duration RF-ablations. The effect of perfusion, and thermal diffusion thereof, can potentially be eliminated by using high intensities with sonication durations less than few seconds \citep{billard1990effects, chen1993effect}. However, due to high losses in the propagation to the kidney, this could lead to significant pre-focal heating and possible skin damage \citep{watkin1996intensity}.

Another phenomenon that has been shown to reduce the efficacy of renal HIFU therapy is respiratory movement \citep{schwartz1994kidney, marberger2005extracorporeal}. The respiration-induced movement of kidneys has been shown to be approximately 16-17 mm in the craniocaudal direction (i.e., from head to feet) \citep{bussels2003respiration}, which is large compared to the radial size of the simulated focal points ($\sim$1.5 mm). This effect was not incorporated in the simulation model, but could potentially result in significant reduction in heating efficiency and generation of unintended lesions caused by overheating of adjacent healthy tissue. In practice, this effect can be controlled using respiratory gating \citep{muller2013management}, but in this case the sonication durations have to be significantly shorter than used here.

\section*{Conclusions}

The efficacy of HIFU therapy of the kidney was investigated with fully three-dimensional acoustic and thermal simulations in three different patients. The acoustic simulations showed that the intensity of the ultrasound field dropped on average by 11.1 dB and it was found that the intensity loss could be roughly divided equally between attenuation and refraction. Reflections due to tissue interfaces were estimated to be less than 0.1 dB and the rib cage was avoided by positioning of the transducer.

A key contribution of this work was quantifying the effect refraction has on: splitting of the focal zone, the thermal dose and shifting of the position of the focus. It was necessary for a 3-D model to quantify these effects as refraction is sensitive to the relative angles of the acoustic beam and sound speed gradients. Refraction resulted in a three-fold drop in peak temperature and a eleven-fold reduction in the ablated volume; but also produced large patient-to-patient variability with one patient having almost no ablation at all. This variability is consistent with that reported in clinical outcomes for kidney tumours. Biological variability of many properties in the kidney was considered and no large differences in temperature elevation were seen with short sonication durations.

The results reported here indicate that focal splitting to be a significant factor affecting the efficacy of HIFU treatment of kidney tumours. Back-propagation simulations suggested that patient-specific phase correction at the source should be able to mitigate the effects of refraction and also minimise patient-to-patient variability.


%

%

\section*{Acknowledgements}

V.~S. acknowledges the support of the RCUK Digital Economy Programme grant number EP/G036861/1 (Oxford Centre for Doctoral Training in Healthcare Innovation) as well as the support of Instrumentarium Science Foundation, Jenny and Antti Wihuri Foundation, Finnish Cultural Foundation, Finnish Foundation for Technology Promotion and Otto A. Malm Foundation. J.~J. is financed from the SoMoPro II Programme, co-financed by the European Union and the South-Moravian Region. This work reflects only the authors' views and the European Union is not liable for any use that may be made of the information contained therein. B.~T. acknowledges the support of EPSRC grant numbers EP/L020262/1 and EP/M011119/1. R.~C. acknowledges the support of EPSRC grant number EP/K02020X/1.





\pagebreak

\bibliographystyle{UMB-elsarticle-harvbib}

\end{document}